\documentclass[prl,onecolumn,superscriptaddress,nofootinbib]{revtex4}

\usepackage[pdftex]{graphicx}

\usepackage{multirow}
\usepackage{url}
\usepackage{enumerate}
\usepackage{enumitem}
\usepackage{amsmath,amssymb,latexsym}
\usepackage{ascmac}
\usepackage{bm}
\def\U#1{{\rm #1}}
\def\tr{\U{tr}}
\def\({\left(}
\def\){\right)}

\newtheorem{theorem}{{\bf Theorem}}
\newtheorem{lemma}{{\bf Lemma}}
\newcommand{\bra}[1]{\langle #1 |}
\newcommand{\ket}[1]{| #1 \rangle}

\begin{document}
\title
{Finite-key security analysis of the decoy-state BB84 QKD with passive measurement}

\author{Akihiro Mizutani}
\affiliation{Faculty of Engineering, University of Toyama, Gofuku 3190, Toyama 930-8555, Japan}
\author{Shun Kawakami}
\affiliation{Network Innovation Laboratories, NTT Inc., 1-1 Hikari-no-oka, Yokosuka, Kanagawa, 239-0847, Japan}
\affiliation{NTT Research Center for Theoretical Quantum Information, NTT Inc., 3-1 Morinosato Wakanomiya, Atsugi, Kanagawa, 243-0198, Japan}
\author{Go Kato}
\affiliation{Quantum ICT Laboratory, NICT, Koganei, Tokyo 184-8795, Japan}

\begin{abstract}
  {
The decoy-state Bennett-Brassard 1984 (BB84) quantum key distribution (QKD) protocol is widely regarded as the de facto standard for practical implementations. On the receiver side, passive basis choice is attractive because it 
significantly reduces the need for random number generators and eliminates the need for optical modulators. 
Despite these advantages, a finite-key analytical security proof for the decoy-state BB84 protocol, where the basis is chosen passively with a biased probability, has been lacking. In this work, we present a simple analytical finite-key security proof for this setting, yielding a closed-form secret-key rate formula that can be directly evaluated using experimentally accessible parameters. Numerical simulations show that the key rates of passive- and active-measurement implementations are nearly identical, indicating that passive measurement does not compromise key-generation efficiency in practical QKD systems.
}
 \end{abstract}

\maketitle

\section{Introduction}
Quantum key distribution (QKD) is a well-established research field in quantum information processing. 
Since the proposal of the Bennett-Brassard 1984 (BB84) protocol~\cite{Bennett1984}, numerous QKD protocols have been developed. 
Among them, the BB84 protocol combined with the decoy-state method~\cite{decoy1, decoy2, decoy3} is widely regarded as the de facto standard for practical and commercial implementations~\cite{Toshiba,NEC,idQ}, owing to its simplicity and well-established security proofs (e.g.,~\cite{Devashish2025, pspd}). 
In the decoy-state method, the sender (Alice) employs weak coherent pulses instead of true single-photon sources without compromising the communication distance. 
On the receiver side (Bob), threshold (on/off) detectors are typically used, and the security of the decoy-state BB84 protocol employing threshold detectors has been proven using the squashing model~\cite{squash1, squash2}, complementarity-based proofs~\cite{complementary1, complementary2}, and the entropic uncertainty principle~\cite{Tomamichel2011}.

In the BB84 protocol, Alice and Bob each choose one of the two bases, and there are two important aspects of this basis choice. First, biasing the basis choice increases the key generation rate~\cite{Lo2004}. Specifically, one basis is used primarily for key generation, whereas the other monitors channel disturbance. Second, the choice can be implemented passively rather than actively. Passive optical setups are attractive in practice because they significantly reduce the need for random number generators and eliminate the need for optical modulators; the latter introduce optical loss and are limited in switching speed. 
Fully passive state-preparation schemes for Alice have been proposed~\cite{fullypassive}, but they are technically demanding because they require replacing multiple active modulators, such as polarization and intensity modulators, with passive components. In contrast, passive measurement schemes for Bob are relatively simple: a beam splitter directs incoming light into two paths, each corresponding to one basis. This configuration has been used in a wide range of QKD demonstrations, from early free-space systems~\cite{Hughes2002, Schmitt2007} to more recent satellite~\cite{Liao2017, Sivasankaran2022, Roger2023, Li2025} and drone-based QKD~\cite{Conrad2025}, 
as well as fiber-based QKD using polarization~\cite{Qin2025, Chalupnik2025} and time-bin encoding~\cite{timebin1, timebin2, timebin3, timebin4, timebin5, timebin6}. 

A major challenge in proving the security of the BB84 protocol with passive measurement and threshold detectors arises when Bob's basis choice is biased. In such passive-biased BB84 protocols, no squashing map exists~\cite{Lars1}, 
and standard proofs based on complementarity or the entropic uncertainty principle, commonly used for active protocols, apply only when the basis choice is unbiased~\cite{Devashish2025, Kawakami2025}. This difficulty stems from the dependence of Bob's basis-choice probability on the number of incoming photons, which leads, for example, to discrepancies in the estimated phase error rate between passive- and active-biased protocols within complementarity-based approaches~\cite{Kawakami2025}. While recent studies have proven the security of the passive-biased protocol using numerical methods~\cite{Lars1, Lars2, Lars3}, those approaches require solving convex optimization problems to obtain key rates, resulting in additional computational overhead. 
In contrast, a recent analytical proof for the passive-biased protocol~\cite{Kawakami2025} provides a closed-form secret-key-rate formula that can be evaluated directly using experimentally accessible parameters. In that approach, Bob's single-photon detection rate is estimated from cross-click events (simultaneous detections in both paths) enabling phase-error-based proofs in the single-photon subspace, but it applies only in the asymptotic regime.

In this work, we present a fully analytical finite-key security proof for the passive decoy-state BB84 protocol, where Bob passively chooses his basis with a biased probability. Unlike~\cite{Kawakami2025}, 
our proof does not require estimating Bob's single-photon detection rate, enabling a simpler and more direct analysis, 
as detailed in Sec.~\ref{sec:comparison}. 
We show that the discrepancy in phase error rates between passive- and active-biased protocols can be upper-bounded by the cross-click rate, which is typically negligible, as shown in Theorem~\ref{theorem:Nphupper}. Our numerical results indicate that the key-rate difference between active and passive basis choice is negligible, confirming that passive measurement schemes maintain efficiency while offering practical implementation advantages.

This paper is organized as follows. Section~\ref{sec:assumptions} summarizes the assumptions regarding Alice's and Bob's devices. 
Section~\ref{sec:actualprotocol} describes the decoy-state BB84 protocol with Bob's passive basis choice. Section~\ref{sec:securityproof} presents the security proof of the protocol based on phase error rate estimation. 
In Section~\ref{sec:simulation}, we present numerical simulation results and compare them with those for the active protocol. Finally, Section~\ref{sec:conclusions} concludes the paper.

\textit{Note added.} 
After submitting our paper to Quantum Science and Technology, we became aware of an independent work that appeared on 
arXiv~\cite{deva2025}, which studies the finite-key security of passive QKD using a different approach from ours.

\section{Technical differences from the existing works}
\label{sec:comparison}
In this section, we explain the technical differences between the present work and the existing studies on passive-biased BB84 protocols~\cite{Lars1, Kawakami2025,Lars2,Lars3}. Among them, Refs.~\cite{Lars1,Lars2,Lars3} rely on numerical optimization methods to evaluate the key rate, which is a different technique from that adopted in this paper. 
On the other hand, Ref.~\cite{Kawakami2025} provides an analytical security proof of this protocol in the asymptotic regime, 
whose analytical framework is similar to that of the present work. 
Therefore, in the following, we focus on a detailed comparison with Ref.~\cite{Kawakami2025}.

We first note that both Ref.~\cite{Kawakami2025} and the present work base their security proofs on the phase-error-correction (PEC) approach~\cite{complementary1, complementary2} (see, for example, Sec.~6.2 of Ref.~\cite{dev:review} for a review), which is one of the major frameworks for proving the security of QKD protocols. 
The key idea in Ref.~\cite{Kawakami2025} is to exploit the fact that the phase error rate, a central quantity in the PEC approach, can be decomposed into contributions from each photon-number subspace $m$ entering Bob's measurement setup. 
By using this property, the authors focused on the single-photon $(m=1)$ reception events, which dominate the sifted-key events, and derived the corresponding phase error rate. 
For the multiphoton ($m\ge2$) reception events, they assumed the worst-case phase error rate of 0.5. 
This treatment effectively reduces the security proof of the passive-measurement setup to that of the active-measurement case, since for single-photon reception events, both setups perform identical measurements.

In short, Ref.~\cite{Kawakami2025} restricts Bob's incoming signals to single-photon states, and the security proof is carried out within the corresponding restricted Hilbert space. 
This method is a specific instance of the flag-state-squasher technique~\cite{yanbao}, which has been employed in all previous analyses of passive-biased BB84 protocols~\cite{Lars1, Kawakami2025,Lars2,Lars3}. 
Specifically, when this technique is applied within the PEC framework, the following two quantities must be determined to obtain the key rate
\footnote{
Note that in Ref.~\cite{Kawakami2025}, Eq.~(63) provides the upper bound on the number of Bob's detections originating from multiphoton receptions, which corresponds to item~\ref{itemi}, whereas Eq.~(30) corresponds to item~\ref{itemii}.}:
\begin{enumerate}[label=(\roman*)]
\item
\label{itemi}
a lower bound on the number of low-photon-number reception events, and
\item
\label{itemii}
the phase error rate associated with those low-photon-number reception events.
\end{enumerate}

We recall that for active-measurement BB84 protocols, it is not necessary to restrict the photon number entering Bob's measurement setup, and security proofs that do not rely on the flag-state-squasher technique have already been well established~(see, for example,~\cite{complementary2,tomamichel2012,concise2014}).
In contrast, for passive-measurement setups, all existing security proofs to date~\cite{Lars1, Kawakami2025,Lars2,Lars3} have relied on the flag-state-squasher technique, which requires an additional estimation for the number of low-photon-number reception events (namely, item~\ref{itemi} above).

In this work, we take a different approach from Ref.~\cite{Kawakami2025} and establish a security proof for the passive-biased BB84 protocol without employing the flag-state-squasher technique, in the finite-key regime.
Specifically, instead of restricting the phase-error estimation to single-photon $(m=1)$ reception events or adopting the worst-case assumption for multiphoton $(m\ge2)$ receptions, we directly bound the overall phase error rate without restricting $m$ by using experimentally observed quantities such as the bit-error rate in the $X$ basis and the cross-click rate 
(the rate of events where detections occur in both the $X$ and $Z$ lines). 
This approach eliminates the need to estimate the number of low-photon-number reception events in Bob's measurement setup (that is, item~\ref{itemi} above) and thus provides a simpler and more direct analytical security proof than that of Ref.~\cite{Kawakami2025}. 
The direct derivation of the overall phase error rate constitutes the core of our security proof and is formally presented in Lemma~\ref{theorem1:probability}, with the detailed proof given in Appendix~\ref{app:proofLemma1}. 
After obtaining Lemma~\ref{theorem1:probability}, the remaining part of our security proof proceeds as follows. 
In Eq.~(\ref{eq:RHSlemma}) of Lemma~\ref{theorem1:probability}
\footnote{
Note that the superscript ``1" in the random variables used in our security proof (such as $E^{\U{ph},1}_i$ in Eq.~(\ref{eq:RHSlemma})) refers to the number of photons emitted by Alice, not to the number of photons entering Bob's measurement setup.
}, the overall phase-error probability on the left-hand side is upper-bounded by the sum of the probabilities of the $X$-basis bit-error event and the cross-click event for different intensities.
By applying Kato's inequality~\cite{Katoinequality}, a probabilistic concentration inequality, we relate this sum of phase-error probabilities to the number of phase errors, as shown in Theorem~\ref{theorem:Nphupper}. 
Finally, by combining the lower bound on the number of detection events originating from Alice's single-photon emissions in the 
$Z$ basis with the decoy-state method, as shown in Theorem~\ref{theorem:decoy1ZL}, we complete the security proof. 

\section{Assumptions on devices}
\label{sec:assumptions}
Before describing the actual QKD protocol, we summarize the assumptions regarding Alice's source and Bob's measurement. 
As described in Sec.~\ref{sec:actualprotocol}, we adopt the decoy-state BB84 protocol, in which, in addition to emitting signal pulses, two decoy states with different intensities are used. For simplicity of our analysis, we consider that the weakest decoy state is the vacuum state. 

\subsection{Assumptions on Alice's source}
First, we list up the assumptions on Alice's source as follows. 
\begin{enumerate}[label=A\arabic*]
\item
\label{Ass:Alicesingle}
For the $i$th pulse emission with $1\le i\le N$, Alice uniformly and randomly chooses bit $a_i\in\{0,1\}$, 
intensity label $\omega_i\in\{S,D,V\}$ with probability $p_{\omega}:=p(\omega_i=\omega)$, and basis 
$\alpha_i\in\{Z,X\}$ with probabilities $p_Z:=p(\alpha_i=Z)$ and $p_X:=p(\alpha_i=X)$. 
Here, $S$, $D$, and $V$ represent the signal, decoy, and vacuum, respectively.
\item
The intensity when the intensity label $\omega_i=\omega$ is chosen is denoted by $\mu_{\omega}$. 
For these intensities, we suppose that $\mu_V=0$ and $0<\mu_D<\mu_S$ hold. 
\item
\label{assA:prcs}
Depending on Alice's choices of $a_i,\omega_i$ and $\alpha_i$, she prepares the following phase-randomized coherent state
\begin{align}
\hat{\rho}_{S_i}(a_i,\omega_i,\alpha_i)
&:=
\frac{1}{2\pi}\int_0^{2\pi}
\ket{\sqrt{\mu_{\omega_i}}e^{\U{i}\delta},\phi(a_i,\alpha_i)}
\bra{\sqrt{\mu_{\omega_i}}e^{\U{i}\delta},\phi(a_i,\alpha_i)}_{S_i}
\U{d}\delta
\notag\\
&=\sum_{n=0}^{\infty}p^{\U{int}}_{\omega_i}(n)
\ket{n_{\phi(a_i,\alpha_i)}}\bra{n_{\phi(a_i,\alpha_i)}}_{S_i}
\label{eq:assAlicesourceCS}
\end{align}
and sends the state of system $S_i$ to Bob via a quantum channel. 
Here, $\ket{n_{\phi(a_i,\alpha_i)}}$ denotes an $n$-photon number state with polarization $\phi(a_i,\alpha_i)$, 
where $\phi(0,Z),\phi(1,Z),\phi(0,X)$ and $\phi(1,X)$ are horizontal, vertical, diagonal and anti-diagonal polarizations, 
respectively. 
The coherent state $\ket{\sqrt{\mu}e^{\U{i}\delta},\phi(a,\alpha)}$ with polarization $\phi(a,\alpha)$ is defined by
\begin{align}
\ket{\sqrt{\mu}e^{\U{i}\delta},\phi(a,\alpha)}:=e^{\frac{-\mu}{2}}\sum_{n=0}^{\infty}\frac{(\sqrt{\mu}e^{\U{i}\delta})^n}{\sqrt{n!}}
\ket{n_{\phi(a,\alpha)}}.
\end{align}
The probability of emitting $n$ photons is written as
\begin{align}
p^{\U{int}}_{\omega}(n)=e^{-\mu_{\omega}}\mu_{\omega}^n/n!.
\end{align}
\end{enumerate}

\subsection{Assumptions on Bob's measurement}
Next, we summarize the assumptions on Bob's measurement as follows. 
\begin{enumerate}[label=B\arabic*]
\item
Bob's measurement setup consists of a beam splitter (BS) and $Z$ and $X$ lines. 
The $Z$ line contains a polarization beam splitter (PBS) followed by two threshold photon detectors, 
while the $X$ line contains a half-wave plate (HWP) and a PBS, also followed by two threshold photon detectors. 
Here, threshold detectors only indicate the arrival of signals and cannot distinguish the number of photons. 
Since we consider a passive measurement setup, Bob's basis $\beta_i\in\{Z,X\}$ for the $i$th incoming pulse is determined by whether a click event occurs only on the $Z$ line or the $X$ line.
\item
\label{Ass:BobBS}
Bob's BS is sensitive only to the number of incoming photons, whose transmittance is denoted by $q$ with $0<q<0.5$. 
\item
\label{Ass:Bobdetector}
The four threshold photon detectors have detection efficiency $\eta_{\U{det}}$ with $0<\eta_{\U{det}}\le1$ 
and the dark count probability $d$ with $0\le d<1$
\footnote{
Note that when the  detection efficiencies are not identical, especially under multiphoton inputs, simple 
post-processing balancing (such as random discarding or weighting of detection events) cannot necessarily justify the security (see for example, \cite{quantummulti}).
}. 
\end{enumerate}

\section{Actual protocol}
\label{sec:actualprotocol}
\begin{figure*}[t]
    \centering
    \includegraphics[width=12.5cm]{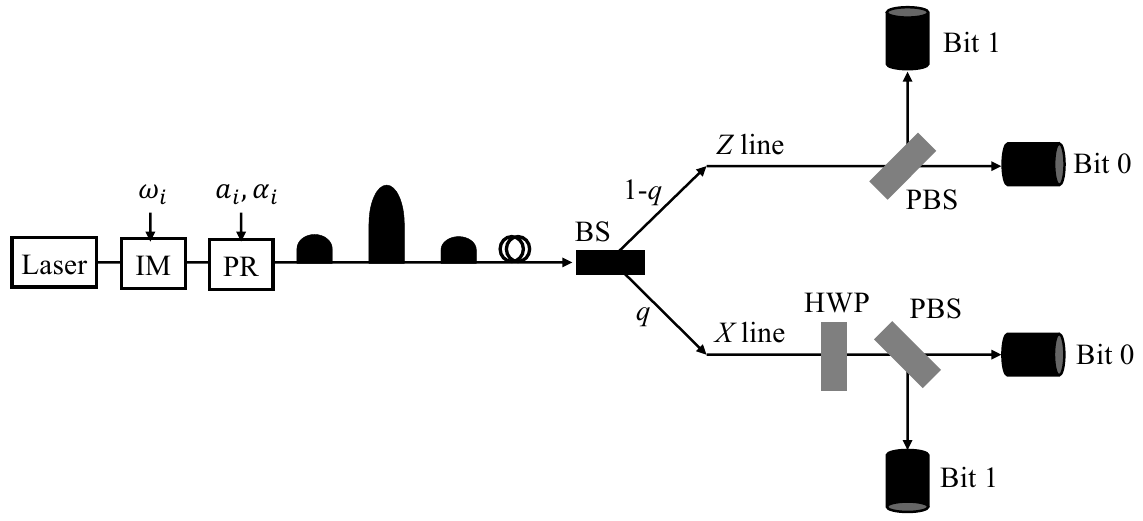}
    \caption{
    Schematic of our passive decoy-state BB84 protocol. For each $i$th emission, Alice sends to Bob a phase randomized coherent state $\hat{\rho}_{S_i}(a_i,\omega_i,\alpha_i)$ with her choices of bit $a_i$, basis $\alpha_i$ and intensity label $\omega_i$. The state is characterized by its intensity $\mu_{\omega_i}$, set by the intensity modulator (IM), and its polarization $\phi(a_i,\alpha_i)$, controlled by the polarization rotator (PR). 
    Bob has a passive measurement setup, where the incoming pulse passes through the beam splitter (BS), whose transmittance is $q$, followed by the $Z$ and $X$ lines. 
    The $Z$ line consists of a polarization beam splitter (PBS) and two threshold photon detectors with detection efficiency $\eta_{\U{det}}$ and dark count probability $d$. The $X$ line consists of a half-wave plate (HWP), a polarization beam splitter (PBS), and two threshold photon detectors with the same detection efficiency and dark count probability. 
    }   
    \label{fig:actual}
\end{figure*}

We describe our decoy-state BB84 protocol with passive measurement (see Fig.~\ref{fig:actual}).
\begin{enumerate}[label=P\arabic*]
\item
Alice and Bob, respectively, repeat the following procedures for $i=1$ to $i=N$. 
\begin{enumerate}
\item
\label{step:A1}
Alice uniformly and randomly chooses bit $a_i$, basis $\alpha_i\in\{Z,X\}$ with probability 
$p_{\alpha_i}$, and the intensity label $\omega_i\in\{S,V,D\}$ with probability $p_{\omega_i}$. 
She sends a phase-randomized coherent state $
\hat{\rho}_{S_i}(a_i,\omega_i,\alpha_i)$ defined in Eq.~(\ref{eq:assAlicesourceCS}) to Bob via a quantum channel.
\item
Bob forwards the incoming pulse to his measurement setup and records which of the four detectors click. 
If no detection occurs in all the detectors, Bob obtains a {\it no-click event}. 
If only one detector clicks on the $Z$ ($X$) line, Bob sets bit $b_i$ to the value of which detector clicked. 
If both detectors click on the $Z$ ($X$) line and no detection occurs on the other line, 
Bob sets $b_i$ to a random bit value. 
If a detection occurs on both lines, Bob obtains a {\it cross-click event}. 
Bob sets the basis choice to $\beta_i=Z$ or $\beta_i=X$ depending on whether he obtains a click event only 
on the $Z$ line or the $X$ line.
\end{enumerate}
\item
\label{actualstep:Bobrecord}
Bob records an index set $\mathcal{D}_Z\subseteq\{1,2,...,N\}$ of click events with $\beta_i=Z$, 
the one $\mathcal{D}_X\subseteq\{1,2,...,N\}$ of click events with $\beta_i=X$, 
and the one $\mathcal{D}_{\U{cross}}\subseteq\{1,2,...,N\}$ of cross-click events. 
Bob obtains the $X$-basis string $\bm{b}_X:=(b_i)_{i\in \mathcal{D}_X}$.
\item
\label{step:bobData}
Bob announces $\mathcal{D}_Z$, $\mathcal{D}_X$, $\mathcal{D}_{\U{cross}}$ and $\bm{b}_X$ 
to Alice via an authenticated classical channel. 
\item
\label{step:DZ'}
Alice defines an index set $\mathcal{D}'_Z:=\{i\in\mathcal{D}_Z|\alpha_i=Z\}$ with length $N_{\U{sift}}:=|\mathcal{D}'_Z|$ 
and announces $(\alpha_i)_{i\in\mathcal{D}'_Z}$ to Bob via an authenticated classical channel. 
\item
Bob obtains his sifted key $\bm{b}_B:=(b_i)_{i\in \mathcal{D}'_Z}$.
\item
\label{step:dataAlice}
Alice obtains her sifted key $\bm{a}_A:=(a_i)_{i\in \mathcal{D}'_Z}$, 
$N_{Z,\omega}:=|\{i\in\mathcal{D}'_Z|\omega_i=\omega\}|$, 
$N^{\U{Error}}_{X,\omega}:=|\{i\in\mathcal{D}_X|\alpha_i=X,\omega_i=\omega,a_i\neq b_i\}|$, 
and $N_{\omega}^{\U{Cross}}:=|\{i\in\mathcal{D}_{\U{cross}}|\omega_i=\omega\}|$. 
Alice calculates the amount of privacy amplification $N_{\U{PA}}$ using $\{N_{Z,\omega}\}_{\omega\in\{S,D,V\}}$, 
$\{N^{\U{Error}}_{X,\omega}\}_{\omega\in\{S,D,V\}}$ and $\{N_{\omega}^{\U{Cross}}\}_{\omega\in\{S,D,V\}}$, 
and sends the value of $N_{\U{PA}}$ to Bob via an authenticated classical channel. 
\item
(Bit error correction) 
Alice chooses and announces a bit error correcting code and sends the syndrome on her sifted key $\bm{a}_A$ 
to Bob by consuming a pre-shared secret key of length $N_{\U{EC}}$. 
Using the syndrome information sent by Alice, Bob corrects the bit errors in his sifted key $\bm{b}_B$ 
and obtains the corrected key. 
\item
(Privacy amplification) 
Alice and Bob execute privacy amplification by shortening their reconciled keys by $N_{\U{PA}}$ bits to 
respectively share the final keys $\bm{k}_A$ and $\bm{k}_B$ of length 
\begin{align}
N_{\U{fin}}:=N_{\U{sift}}-N_{\U{PA}}.
\end{align}
\end{enumerate}
By executing this protocol, the bit size of the increased secret key  is given by
\begin{align}
\ell=N_{\U{sift}}-N_{\U{PA}}-N_{\U{EC}}.
\label{eq:eqell}
\end{align}

\section{SECURITY PROOF}
\label{sec:securityproof}
In this section, we provide a finite-key security analysis for the QKD protocol described in Sec.~\ref{sec:actualprotocol}. 
Below, we summarize the definitions that we use throughout this paper. 
\begin{enumerate}
\item
The trace norm for linear operator $\hat{A}$ is defined by 
$||\hat{A}||:=\tr\sqrt{\hat{A}^{\dagger}\hat{A}}$. 
\item
Let $\ket{0}$ and $\ket{1}$ denote the $Z$-basis states, and 
$\ket{+}$ and $\ket{-}$ denote the $X$-basis states with $\ket{\pm}=(\ket{0}\pm\ket{1})/\sqrt{2}$. 
\item
For set $\Omega$, its cardinality is denoted by $|\Omega|$.
\item
A function $h(x)$ is defined as
\begin{align}
h(x)=
\begin{cases}
-x\log_2x-(1-x)\log_2(1-x)&0\le x\le0.5\\
1&0.5<x.
\end{cases}
\label{eq:binary}
\end{align}
\item
$\delta(x,y)$ denotes the Kronecker delta, which equals 1 if $x=y$, 0 otherwise.
\end{enumerate}

\subsection{Security criteria and secret key rate}
Here, we explain the security criteria that we adopt in our security analysis. 

In so doing, we introduce state $\hat{\rho}^{\U{actual}}_{AE|N_{\U{fin}}}$ of Alice's final key and Eve's quantum 
system after executing the actual protocol: 
\begin{align}
\hat{\rho}^{\U{actual}}_{AE|N_{\U{fin}}}:=
\sum_{\bm{k}_A\in\{0,1\}^{N_{\U{fin}}}}P(\bm{k}_A|N_{\U{fin}})\ket{\bm{k}_A}\bra{\bm{k}_A}_A\otimes
\hat{\rho}^{\U{actual}}_{E|N_{\U{fin}}}(\bm{k}_A).
\end{align}
Here, $P(\bm{k}_A|N_{\U{fin}})$ denotes the probability of Alice obtaining the final key $\bm{k}_A$ 
when the final key length is $N_{\U{fin}}$, and $\hat{\rho}^{\U{actual}}_{E|N_{\U{fin}}}(\bm{k}_A)$ denotes Eve's state when Alice's final key is $\bm{k}_A$ and the key length is $N_{\U{fin}}$. 

Also, we introduce the ideal state, where Alice's final key is completely random and independent of Eve's system when the 
final key length is $N_{\U{fin}}$: 
\begin{align}
\hat{\rho}^{\U{ideal}}_{AE|N_{\U{fin}}}:=\frac{1}{2^{N_{\U{fin}}}}
\sum_{\bm{k}\in\{0,1\}^{N_{\U{fin}}}}\ket{\bm{k}}\bra{\bm{k}}_A\otimes\tr_A[\hat{\rho}^{\U{actual}}_{AE|N_{\U{fin}}}].
\end{align}
In this work, we adopt the universal composable security criteria~\cite{composable1,composable2}, which is the most widely used criteria in QKD. 
In this framework, the QKD protocol is $\epsilon_{\U{sec}}$-secure if the protocol is $\epsilon_c$-correct and $\epsilon_s$-secret with $\epsilon_{\U{sec}}=\epsilon_c+\epsilon_s$. 
Here, the protocol is $\epsilon_c$-correct if the probability of obtaining different final keys is upper-bounded by $\epsilon_c$, namely, $P(\bm{k}_A\neq\bm{k}_B)\le\epsilon_c$. 
Also, the protocol is $\epsilon_s$-secret if 
\begin{align}
\frac{1}{2}\sum_{N_{\U{fin}}=0}^NP(N_{\U{fin}})
||\hat{\rho}^{\U{actual}}_{AE|N_{\U{fin}}}-\hat{\rho}^{\U{ideal}}_{AE|N_{\U{fin}}}||\le \epsilon_s
\label{eq:defsecrecy}
\end{align}
holds. Here, $P(N_{\U{fin}})$ denotes the probability of obtaining the final key of length $N_{\U{fin}}$ in the actual protocol. 

There are various methods to derive the secrecy parameter $\epsilon_s$, such as the phase error correction approach~\cite{complementary1,complementary2}, which we adopt in this work. In this approach, Step~\ref{step:A1} of Alice's procedure in the actual protocol is replaced by a step employing an entangled state. Specifically, instead of Alice choosing $(a_i,\omega_i,\alpha_i)$ and emitting state $\hat{\rho}_{S_i}(a_i,\omega_i,\alpha_i)$, Alice prepares the following purified state of Eq.~(\ref{eq:assAlicesourceCS}):
\begin{align}
\sum_{a_i\in\{0,1\}}\sum_{\omega_i\in\{S,D,V\}}
\sum_{n_i=0}^{\infty}\sum_{\alpha_i\in\{Z,X\}}\sqrt{p_{\omega_i}p^{\U{int}}_{\omega_i}(n_i)p_{\alpha_i}/2}
\ket{a_i}_{A_i^{\U{bit}}}\ket{\omega_i}_{A_i^{\U{amp}}}\ket{n_i}_{N_i}\ket{\alpha_i}_{A_i^{\U{basis}}}
\ket{{n_{i,\phi(a_i,\alpha_i)}}}_{S_i}
\label{eq:alicestatevirtual}
\end{align}
and sends system $S_i$ to Bob. Then she measures systems $A_i^{\U{bit}},A_i^{\U{amp}},A_i^{\U{basis}}$ to obtain $(a_i,\omega_i,\alpha_i)$. Here, $n_i\in\{0,1,2,...,\}$ denotes the number of photons contained in the $i$th emitted pulse.

For the events where Bob obtains a click only on the $Z$ line with $\alpha_i=Z$ (with the corresponding index set $\mathcal{D}'_Z$ defined in Step~\ref{step:DZ'}), we assume that Alice's virtual $X$-basis measurement outcomes $\bm{k}_{\U{ph}}\in\{+,-\}^{|\mathcal{D}'_Z|}$ on systems 
$A_i^{\U{bit}}$ for $i\in\mathcal{D}'_Z$ are contained within a certain set $\Omega_{\U{ph}}\subset\{+,-\}^{|\mathcal{D}'_Z|}$ with probability at least $1-\epsilon_{\U{ph}}$; that is, 
\begin{align}
P(\bm{k}_{\U{ph}}\in\Omega_{\U{ph}})\ge1-\epsilon_{\U{ph}}
\label{eq:kphprobability}
\end{align}
holds regardless of Eve's attack. Then, by setting the amount of privacy amplification as
\begin{align}
N_{\U{PA}}=\log_2|\Omega_{\U{ph}}|+\xi,
\label{eq:NPA}
\end{align}
for $\xi>0$, the protocol is $\epsilon_s$-secret with $\epsilon_s=\sqrt{2}\sqrt{\epsilon_{\U{ph}}+2^{-\xi}}$~\cite{complementary2,HayashiTsurumaru2012}.
In the phase error correction approach, we consider correcting Alice's qubits of systems $A_i^{\U{bit}}$ for 
$i\in\mathcal{D}'_Z$ to an eigenstate $\ket{+}^{\otimes |\mathcal{D}'_Z|}$ of the 
$X$ basis (a complementary basis of the key generation basis). 
The amount of phase error syndrome required to correct Alice's qubits to the eigenstate $\ket{+}^{\otimes |\mathcal{D}'_Z|}$ is 
$N_{\U{PA}}$. Hence, the smaller $|\Omega_{\U{ph}}|$ is, the fewer sifted key bits need to be sacrificed for syndrome measurement.  Importantly, $|\Omega_{\U{ph}}|$ cannot be directly obtained in the actual protocol and must instead be estimated from the data obtained in the actual protocol. 

\subsection{Derivation of the number of Alice's virtual $X$-basis measurement outcomes $|\Omega_{\U{ph}}|$}
\label{subsec:vscenario}
Our remaining task in the security proof is to derive the upper bound on $|\Omega_{\U{ph}}|$ using experimentally available data. 
A key point in the phase error correction
approach is that, when deriving the secrecy parameter of Eq.~(\ref{eq:defsecrecy}), 
Bob's system $B$ is traced out. Therefore, Bob is allowed to perform any measurement to improve the estimation accuracy of Alice's virtual $X$-basis measurement outcomes $\bm{k}_{\U{ph}}\in\{+,-\}^{|\mathcal{D}'_Z|}$. In our analysis, we consider that Bob performs the same measurement as on the $X$ line (i.e., $X$-basis measurement) even for the pulses on the $Z$ line.

Keeping this point in mind, we introduce the following virtual protocol for deriving the upper bound on $|\Omega_{\U{ph}}|$, 
which differs from the actual protocol in the following three aspects~\ref{itemD1}-\ref{itemD3}.

\begin{enumerate}[label=D\arabic*]
\item
\label{itemD1}
Alice prepares the state of Eq.~(\ref{eq:alicestatevirtual}) for each pulse emission and sends system $S_i$ to Bob via a quantum channel. 
From Assumption~\ref{assA:prcs}, this virtual operation is indistinguishable from the transmission of system $S_i$ in the state described by Eq.~(\ref{eq:assAlicesourceCS}) in the actual protocol from Eve's perspective. Therefore, this replacement is justified in a virtual protocol.
\item
\label{itemD2}
A beam splitter with transmittance $\eta_{\U{det}}$ is placed immediately before Bob's measurement setup (i.e., at the end of the quantum channel), and measurements on both the $X$- and $Z$-lines are then performed using threshold detectors with unit detection efficiency. This virtual measurement is equivalent to using threshold detectors with non-unit detection efficiency 
$\eta_{\U{det}}$
\footnote{
Note that time-bin phase coding is also a widely adopted passive-basis measurement scheme, but the present analysis for polarization coding cannot be directly applied to the time-bin case. This is because, the $X$ line in a time-bin setup involves three time slots, and only the middle slot, where interference occurs, is used for $X$-basis bit generation (the first and third slots are discarded). Consequently, when the same optical signal enters the $Z$ and $X$ lines, the effective detection efficiency of the $X$ line becomes half that of the $Z$ line. Therefore, it is not possible to represent both measurements by inserting a single beam splitter with a common transmittance $\eta_{\U{det}}$ in front of Bob's measurement setup, as was done in the polarization-based analysis.
}. This modification is justified by Assumption~\ref{Ass:Bobdetector}, which states that all detectors have the same detection efficiency; see, e.g., Theorem~1 of Ref.~\cite{QST2024} for a proof of the equivalence.
\item
\label{itemD3}
Bob performs a quantum nondemolition measurement in the photon-number basis immediately before his beam splitter to obtain the outcome $m_i\in\{0,1,2,...\}$. Subsequently, he performs a measurement using the setup on the $X$ line, regardless of which line the pulse is on. This photon-number measurement is allowed as it is commutable with photon detection by Bob. 
\end{enumerate}

In this  virtual protocol, we consider the following stochastic trials. 
\begin{enumerate}[label=V\arabic*]
\item
\label{step:measuren}
For each pulse emission, Alice measures her systems $A_i^{\U{basis}}$, $A_i^{\U{amp}}$ and $N_i$ 
to obtain the outcomes $\alpha_i$, $\omega_i$ and $n_i$, respectively. 
Although Alice does not know the value of $n_i$ in the actual protocol, this value can be known in principle thanks to Assumption~\ref{assA:prcs}. 
\item
For each pulse reception, Bob measures the number of incoming photons and obtains $m_i\in\{0,1,2,...\}$, 
and Alice and Bob determine the value of the random variable $e_{X,i}\in\{0,1,\emptyset,\U{cross}\}$ 
based on which type of detection occurs. 
\begin{enumerate}
\item
When no detection occurs in all the four detectors, Bob assigns $e_{X,i}=\emptyset$.
\item
\label{step:ZXvir}
When Bob obtains a click event on the $Z$ line and no detection occurs on the $X$ line, 
Bob assigns $\beta_i=Z$. 
When this event occurs with $\alpha_i=Z$, Alice and Bob compare their $X$-basis measurement outcomes and assign 
$e_{X,i}=0$ or 1, 
depending on whether these outcomes are identical or not, respectively~\footnote{Note that, as in the actual protocol, when both detectors click on the $Z$-line or $X$-line, a random bit value is assigned as the measurement outcome.
}. 
Recall that Alice measures system $A_i^{\U{bit}}$ in Eq.~(\ref{eq:alicestatevirtual}) in the $X$ basis, and 
Bob measures the pulses on the $Z$ line using the same measurement setup as for the $X$ line. 
The random variable that corresponds to phase error detection (namely, Alice and Bob obtain different $X$-basis measurement outcomes with $\alpha_i=\beta_i=Z$) when Alice emits $n$ photons in the $i$th round is defined by
\begin{align}
E^{\U{ph},n}_i:=\delta(\alpha_i,Z)\delta(n_i,n)\delta(\beta_i,Z)\delta(e_{X,i},1).
\label{eq:Enphi}
\end{align}
Also, we define the random variables corresponding to the events where $\alpha_i=\beta_i=Z$ in the $i$th round, given that Alice emits $n$ photons and the intensity label is $\omega\in\{S,D,V\}$, as
\begin{align}
Z^{\U{det},n}_i:=
\delta(\alpha_i,Z)\delta(n_i,n)\delta(\beta_i,Z)[\delta(e_{X,i},0)+\delta(e_{X,i},1)]
\label{Zndeti}
\end{align}
and
\begin{align}
Z^{\U{det},\omega}_i:=
\delta(\alpha_i,Z)\delta(\omega_i,\omega)\delta(\beta_i,Z)[\delta(e_{X,i},0)+\delta(e_{X,i},1)],
\label{Zomegadeti}
\end{align}
respectively. 
\\
\item
When Bob obtains a click event on the $X$ line and no detection occurs on the $Z$ line, 
Bob assigns $\beta_i=X$. 
When this event occurs with $\alpha_i=X$, Alice and Bob compare their $X$-basis measurement outcomes and assign 
$e_{X,i}=0$ or 1 depending on whether these outcomes are identical or not, respectively. 
We then define the random variables 
\begin{align}
E^{X,n}_i:=\delta(\alpha_i,X)\delta(n_i,n)\delta(\beta_i,X)\delta(e_{X,i},1)
\end{align}
and
\begin{align}
E^{X,\omega}_i:=\delta(\alpha_i,X)\delta(\omega_i,\omega)\delta(\beta_i,X)\delta(e_{X,i},1)
\label{eq:EXNEXOMEGA}
\end{align}
to represent the occurrence of a bit error in the $X$-basis when Alice emits $n$ photons and the intensity label is 
$\omega\in\{S,D,V\}$, respectively.
\item
When a cross-click event occurs, Bob assigns $e_{X,i}=\U{cross}$. 
The random variables corresponding to this event, when Alice emits $n$ photons and the intensity label is $\omega\in\{S,D,V\}$ 
in the $i$th round, are respectively defined by
\begin{align}
C^n_i:=\delta(n_i,n)\delta(e_{X,i},\U{cross})
\end{align}
and
\begin{align}
C^{\omega}_i:=\delta(\omega_i,\omega)\delta(e_{X,i},\U{cross}).
\label{eq:CnComega}
\end{align} 
\end{enumerate}
\end{enumerate}
The phase error event occurs when Alice and Bob obtain the outcome $e_{X,i}=1$ in Step~\ref{step:ZXvir}. 
We can divide the occurrence of this event according to the number of photons $n_i$ emitted by Alice, which 
is measured in Step~\ref{step:measuren}. 
In our analysis, we consider a worst case scenario where the number of phase error events takes the maximum values 
except for the single-photon emission events. Let 
\[
N_{\U{ph}}^1:=\sum_{i=1}^NE_i^{\U{ph},1}
\]
denote the number of phase error events when Alice emits a single photon in the virtual protocol. Under this worst-case scenario, the number of Alice's 
virtual $X$-basis measurement outcomes with $\alpha_i=\beta_i=Z$, after performing the above virtual protocol for $i=1$ to $N$, is upper bounded as
\begin{align}
|\Omega_{\U{ph}}|\le2^{N_{\U{sift}}-N_Z^1+N_Z^1h\(\frac{N^1_{\U{ph}}}{N_Z^1}\)}.
\label{eq:omega}
\end{align}
Here, $N_{\U{sift}}=|\mathcal{D}'_Z|$ is defined in Step~\ref{step:DZ'}, 
$N_Z^1:=\sum_{i=1}^NZ_i^{\U{det},1}$, and $h(x)$ is defined by Eq.~(\ref{eq:binary}). 
Once we obtain the upper bound on $N^1_{\U{ph}}$ and the lower bound on $N_Z^1$, substituting these bounds to Eq.~(\ref{eq:omega}) gives the amount of privacy amplification $N_{\U{PA}}$ from Eq.~(\ref{eq:NPA}). 
As for the upper bound on $N^1_{\U{ph}}$, we obtain the following Theorem~\ref{theorem:Nphupper}, 
which is our main contribution of this paper. Regarding the lower bound on $N_Z^1$, we have the following Theorem~\ref{theorem:decoy1ZL}, which can be derived from the decoy-state method~\cite{Ma2005}.

In our finite-key analysis, we use a concentration inequality known as Kato's inequality~\cite{Katoinequality,tightnpj} 
(see also Appendix~\ref{app:kato}). 
In the following descriptions of Theorems~\ref{theorem:Nphupper} and \ref{theorem:decoy1ZL},  
\begin{align}
\Delta_{\U{U}}(M,\tilde{M},N,\epsilon):=
\frac{b_{\U{U}}\(N,\tilde{M},\epsilon\)N+a_{\U{U}}\(N,\tilde{M},\epsilon\)(2M-N)
}{\sqrt{N}}\ge0
\label{eq:deviationfunc}
\end{align}
denotes the deviation term of Kato's inequality for upper-bounding the sum of probabilities. Also, 
\begin{align}
\Delta_{\U{L}}(M,\tilde{M},N,\epsilon):=
\frac{b_{\U{L}}\(N,\tilde{M},\epsilon\)N+ a_{\U{L}}\(N,\tilde{M},\epsilon\)(2M-N)
}{\sqrt{N}}\ge0
\label{eq:deviationfunc2}
\end{align}
denotes the deviation term of Kato's inequality for lower-bounding the sum of probabilities. 
Here, each of the four functions $a_{\U{U}}\(N,\tilde{M},\epsilon\)$, $b_{\U{U}}\(N,\tilde{M},\epsilon\)$, $a_{\U{L}}\(N,\tilde{M},\epsilon\)$ and $b_{\U{L}}\(N,\tilde{M},\epsilon\)$ is determined by the number of emitted pulses $N$, 
an estimate $\tilde{M}$ (a constant value before executing the QKD protocol) of the quantity $M$, 
and the failure probability $\epsilon$ associated with Kato's inequality. 
The values of $a_{\U{U}}\(N,\tilde{M},\epsilon\)$ and $a_{\U{L}}\(N,\tilde{M},\epsilon\)$ are real numbers, 
whereas $b_{\U{U}}\(N,\tilde{M},\epsilon\)$ and $b_{\U{L}}\(N,\tilde{M},\epsilon\)$ are positive real numbers satisfying 
$b_{\U{U}}\(N,\tilde{M},\epsilon\)\ge\left|a_{\U{U}}\(N,\tilde{M},\epsilon\)\right|$ and $b_{\U{L}}\(N,\tilde{M},\epsilon\)\ge\left|a_{\U{L}}\(N,\tilde{M},\epsilon\)\right|$. 
The explicit expressions of these functions, as well as the description of Kato's inequality, are provided in Appendix~\ref{app:kato}.

\begin{theorem}
\label{theorem:Nphupper}
The number of phase error events when Bob obtains $\beta_i=Z$ and Alice emits a single photon in the $Z$ basis, 
and Alice and Bob obtain different outcomes in their $X$-basis measurements, is upper-bounded as follows except with 
probability $5\epsilon$ with $0<\epsilon<1$. 
\begin{align}
&N_{\U{ph}}^1
\le 
\left(1-\frac{2a_{\U{L}}(N,\tilde{N}_{\U{ph}}^1,\epsilon)}{\sqrt{N}}\right)^{-1}
\notag\\
&\Bigg\{
\frac{p^{\U{int}}(1)}{\mu_D}\Bigg[c_1\(\frac{e^{\mu_D}
(N_{X,D}^{\U{Error}}+\Delta_{\U{U}}(N_{X,D}^{\U{Error}},\tilde{N}_{X,D}^{\U{Error}},N,\epsilon))
}{p_D}
-\frac{
N_{X,V}^{\U{Error}}-\Delta_{\U{L}}(N_{X,V}^{\U{Error}},\tilde{N}_{X,V}^{\U{Error}},N,\epsilon)
}{p_V}\)
\notag\\
&+c_2\(\frac{e^{\mu_D}
(N_{D}^{\U{Cross}}+\Delta_{\U{U}}(N_{D}^{\U{Cross}},\tilde{N}_D^{\U{Cross}},N,\epsilon))
}{p_D}
-\frac{
N_{V}^{\U{Cross}}-\Delta_{\U{L}}(N_{V}^{\U{Cross}},\tilde{N}_V^{\U{Cross}},N,\epsilon)
}{p_V}\)\Bigg]
\notag\\
&
+\left(b_{\U{L}}(N,\tilde{N}_{\U{ph}}^1,\epsilon)-a_{\U{L}}(N,\tilde{N}_{\U{ph}}^1,\epsilon)\right)\sqrt{N}
\Bigg\}
\notag\\
&=:N_{\U{ph}}^{1,\U{U}}
\label{eq:theoremNph}
\end{align}
if $a_{\U{L}}(N,\tilde{N}_{\U{ph}}^1,\epsilon)<\frac{\sqrt{N}}{2}$. 
On the other hand, if $a_{\U{L}}(N,\tilde{N}_{\U{ph}}^1,\epsilon)\ge\frac{\sqrt{N}}{2}$, $N_{\U{ph}}^1\le N=:N_{\U{ph}}^{1,\U{U}}$. Here, $p^{\U{int}}(n):=\sum_{\omega\in\{S,D,V\}}p_{\omega}p^{\U{int}}_{\omega}(n)$ denotes the 
probability of Alice emitting $n$ photons in each round, constants $c_1\ge0$ and $c_2\ge0$ are respectively defined by
\begin{align}
c_1:=\frac{1-q}{q}\frac{p_Z}{p_X}
~\U{and}~c_2:=\frac{p_Z(1-q)(1-2q)(1-d)^2}{1-\left(q^2+(1-q)^2\right)(1-d)^2}.
\label{eq:defc1c2}
\end{align}
Note that the values of random variables $\{N_{X,\omega}^{\U{Error}}, N_{\omega}^{\U{Cross}}\}_{\omega\in\{D,V\}}$ can be obtained in Step~\ref{step:dataAlice} of the actual protocol. The quantities $\tilde{N}_{\U{ph}}^1$, $\tilde{N}_{X,\omega}^{\U{Error}}$, and $\tilde{N}_{\omega}^{\U{Cross}}$ are the estimated values of $N_{\U{ph}}^1$, $N_{X,\omega}^{\U{Error}}$, and $N_{\omega}^{\U{Cross}}$, respectively. 
\end{theorem}
We observe that, if the second term (the one multiplied by $c_2$) is removed from $N_{\U{ph}}^{1,\U{U}}$ in Eq.~(\ref{eq:theoremNph}), the number of phase errors equals that of the decoy-state BB84 protocol with active measurement~\cite{pspd}. This implies that the difference between the active- and passive-measurement protocols is 
characterized by the number of cross-click events.

\begin{theorem}
\label{theorem:decoy1ZL}
The number of events where Bob obtains the $Z$ basis ($\beta_i=Z$) when Alice emits a single photon in the $Z$ basis is lower-bounded as follows except with probability $4\epsilon$ with $0<\epsilon<1$. 
\begin{align}
N_{Z}^1\ge
&
\left(1+\frac{2a_{\U{U}}(N,\tilde{N}_{Z}^1,\epsilon)}{\sqrt{N}}\right)^{-1}
\Bigg\{
-\frac{p^{\U{int}}(1)}{p_S}\frac{\mu_De^{\mu_S}}{\mu_S(\mu_S-\mu_D)}
\left[N_{Z,S}+\Delta_{\U{U}}(N_{Z,S},\tilde{N}_{Z,S},N,\epsilon)\right]
\notag\\
&
+\frac{p^{\U{int}}(1)}{p_D}\frac{\mu_Se^{\mu_D}}{\mu_D(\mu_S-\mu_D)}
\left[N_{Z,D}-\Delta_{\U{L}}(N_{Z,D},\tilde{N}_{Z,D},N,\epsilon)\right]
\notag\\
&
-\frac{p^{\U{int}}(1)}{p_V}\frac{\mu_S}{\mu_D(\mu_S-\mu_D)}
\left[N_{Z,V}+\Delta_{\U{U}}(N_{Z,V},\tilde{N}_{Z,V},N,\epsilon)\right]
-\left(b_{\U{U}}(N,\tilde{N}_{Z}^1,\epsilon)-a_{\U{U}}(N,\tilde{N}_{Z}^1,\epsilon)\right)\sqrt{N}
\Bigg\}
\notag\\
=:&N_Z^{1,\U{L}}
\end{align}
if $a_{\U{U}}(N,\tilde{N}_{Z}^1,\epsilon)>-\frac{\sqrt{N}}{2}$. On the other hand, if $a_{\U{U}}(N,\tilde{N}_{Z}^1,\epsilon)\le-\frac{\sqrt{N}}{2}$, $N_{Z}^1\ge0=:N_Z^{1,\U{L}}$. Here, the values of random variables $\{N_{Z,\omega}\}_{\omega\in\{S,D,V\}}$ can be obtained in Step~\ref{step:dataAlice} of the actual protocol. The quantities $\tilde{N}_{Z,\omega}$ and $\tilde{N}_Z^1$ are the estimated values of $N_{Z,\omega}$ and $N_Z^1$, 
respectively. 
\end{theorem}

\subsubsection{Proof of Theorem~\ref{theorem:Nphupper}}
Here, we prove that the number of phase-error events when Alice emits a single photon is upper-bounded as in Eq.~(\ref{eq:theoremNph}). We begin by upper-bounding the probability that a phase-error event occurs and Alice emits a single photon, i.e., the probability that $E^{\U{ph},1}_i$ in Eq.~(\ref{eq:Enphi}) equals 1. We derive the upper bound by using probabilities of events observable in the actual protocol.
The following lemma formalizes this fact.
\begin{lemma}
Let $F_{i-1}$ denote the collection of outcomes $\alpha_i,\omega_i,n_i,m_i,\beta_i,e_{X,i}$ up to round $i-1$. 
Then, for any $F_{i-1}$, the probability of obtaining the phase error event when Alice emits a single photon in the $i$th round is upper-bounded as
\begin{align}
P(E^{\U{ph},1}_i=1|F_{i-1})\le\frac{p^{\U{int}}(1)}{\mu_D}
\Bigg[
&c_1\(\frac{e^{\mu_D}}{p_D}P(E^{X,D}_i=1|F_{i-1})-\frac{P(E^{X,V}_i=1|F_{i-1})}{p_V}\)
\notag\\
+&c_2\(\frac{e^{\mu_D}}{p_D}P(C^{D}_i=1|F_{i-1})-\frac{P(C^{V}_i=1|F_{i-1})}{p_V}\)\Bigg].
\label{eq:RHSlemma}
\end{align}
Note that random variables $E^{X,\omega}_i$ and $C^{\omega}_i$ with $\omega\in\{S,D,V\}$ are defined in Eqs.~(\ref{eq:EXNEXOMEGA}) and (\ref{eq:CnComega}), respectively, and constants $c_1$ and $c_2$ are defined in Eq.~(\ref{eq:defc1c2}).
\label{theorem1:probability}
\end{lemma}
The proof of this lemma is given in Appendix~\ref{app:proofLemma1}. 
Here, the right-hand side of Eq.~(\ref{eq:RHSlemma}) represents the sum of probabilities associated with events observable in the actual protocol. Therefore, if these probabilities can be evaluated using the corresponding random variables, $N_{\U{ph}}^1$ can be upper-bounded using the observed quantities. In performing this evaluation, we apply Kato's inequality, which can be applied to dependent random variables. Note that before inventing Kato's inequality, Azuma's inequality~\cite{azumaineq} was commonly used to prove information-theoretic security of QKD. 
However, since Kato's inequality provides a tighter relation between probabilities and the corresponding random variables, 
it has been widely used in recent finite-key analyses of QKD~\cite{trojan2022,mizutani2023,matsuura2023,cow2024,pspd}. 

As a concrete example of how Kato's inequality is used to relate probabilities to the corresponding random variables, we present a discussion on how $\sum_{i=1}^N P(E^{X,D}_i=1|F_{i-1})$, which is obtained by summing both sides of Eq.~(\ref{eq:RHSlemma}) over $i=1,...,N$, can be bounded by the corresponding random variable $N_{X,D}^{\U{Error}}=\sum_{i=1}^N E^{X,D}_i$. 
For this, Kato's inequality in Eq.~(\ref{eq:katoproupper}) states that
\begin{align}
P\Bigg\{
\sum_{i=1}^NP(E^{X,D}_i=1|F_{i-1})\ge N_{X,D}^{\U{Error}}+
\frac{bN+a\left(2N_{X,D}^{\U{Error}}-N\right)}{\sqrt{N}}
\Bigg\}
\le 
\exp\(-\frac{2b^2-2a^2}{(1+\frac{4a}{3\sqrt{N}})^2}\)
\label{eq:epavirverify}
\end{align}
holds for any constants $a,b\in \mathbb{R}$ satisfying $b\ge|a|$. 
By setting the probability on the right-hand side of Eq.~(\ref{eq:epavirverify}) to a constant $\epsilon$, 
the optimal values of $a$ and $b$ that minimize the deviation term 
$[bN + a(2C - N)] / \sqrt{N}$ for a fixed constant $0 \le C \le N$ 
are given by $a_{\U{U}}(N, C, \epsilon)$ in Eq.~(\ref{kato-ineq-def-a}) and $b_{\U{U}}(N, C, \epsilon)$ in Eq.~(\ref{kato-ineq-def-b}), respectively~(see Sec.~\ref{app:secupper} in Appendix~\ref{app:kato} for their derivations). 
Note that since $N_{X,D}^{\U{Error}}$ is a random variable rather than a constant, it must be fixed to obtain optimal values of $a$ and $b$. To this end, we use an estimated value $\tilde{N}_{X,D}^{\U{Error}}$ of $N_{X,D}^{\U{Error}}$, which can be determined, for example, by pre-testing the QKD protocol prior to its actual execution. 
Using this estimated value $\tilde{N}_{X,D}^{\U{Error}}$, we set 
$a=a_{\U{U}}\(N,\tilde{N}_{X,D}^{\U{Error}},\epsilon\)$ and $b=b_{\U{U}}\(N,\tilde{N}_{X,D}^{\U{Error}},\epsilon\)$ in Eq.~(\ref{eq:epavirverify})
and have that
\begin{align}
\sum_{i=1}^N P(E^{X,D}_i=1|F_{i-1})
&\le
N_{X,D}^{\U{Error}}+
\Delta_{\U{U}}(N_{X,D}^{\U{Error}},\tilde{N}_{X,D}^{\U{Error}},N,\epsilon)
\label{eq:katoNXDerror}
\end{align}
holds except with probability $\epsilon$. Here, the function $\Delta_{\U{U}}$ is defined by Eq.~(\ref{eq:deviationfunc}). 

By applying Kato's inequality in the same manner to the sums of the remaining four conditional probabilities: $\sum_{i=1}^NP(E^{\U{ph},1}_i=1|F_{i-1}), \sum_{i=1}^NP(E^{X,V}_i=1|F_{i-1}), \sum_{i=1}^NP(C^{D}_i=1|F_{i-1})$, and 
$\sum_{i=1}^NP(C^{V}_i=1|F_{i-1})$, which are obtained by summing Eq.~(\ref{eq:RHSlemma}) over $i=1,...,N$, we obtain Theorem~\ref{theorem:Nphupper}. The details are given in Appendix~\ref{app:applKato}.

\subsubsection{Proof of Theorem~\ref{theorem:decoy1ZL}}
Here, we prove Theorem~\ref{theorem:decoy1ZL}. 
From the well-known decoy-state method~\cite{Ma2005}, particularly Proposition~6 in Ref.~\cite{pspd}, the following linear relationship among the probabilities can be obtained as
\begin{align}
P(Z^{\U{det},1}_i=1|F_{i-1})
\ge\sum_{\omega\in\{S,D,V\}}t_{\omega}
P(Z^{\U{det},\omega}_i=1|F_{i-1})
\label{eq:lineardecoyS1}
 \end{align}
with
 \begin{align}
 t_S:=-\frac{p^{\U{int}}(1)}{p_S}\frac{\mu_De^{\mu_S}}{\mu_S(\mu_S-\mu_D)}
 \le0,~~~t_D:=\frac{p^{\U{int}}(1)}{p_D}\frac{\mu_Se^{\mu_D}}{\mu_D(\mu_S-\mu_D)}\ge0,~~~
t_V:=-\frac{p^{\U{int}}(1)}{p_V}\frac{\mu_S}{\mu_D(\mu_S-\mu_D)}
\le0.
\end{align}
Recall that $Z^{\U{det},1}_i$ and $Z^{\U{det},\omega}_i$ with $\omega\in\{S,D,V\}$ are defined in Eqs.~(\ref{Zndeti}) and (\ref{Zomegadeti}), respectively.

If $a_{1,Z}:=a_{\U{U}}(N,\tilde{N}_{Z}^1,\epsilon)>-\frac{\sqrt{N}}{2}$, Kato's inequality in Eq.~(\ref{eq:katoproupper}) states that
\begin{align}
N_{Z}^1\ge\left(1+\frac{2a_{1,Z}}{\sqrt{N}}\right)^{-1}
\left[
\sum_{i=1}^NP(Z^{\U{det},1}_i=1|F_{i-1})-(b_{1,Z}-a_{1,Z})\sqrt{N}
\right]
\label{eq:KatodecoyZ1}
\end{align}
holds except with probability $\epsilon$, where $b_{1,Z}:=b_{\U{U}}(N,\tilde{N}_{Z}^1,\epsilon)$. On the other hand, if $a_{\U{U}}(N,\tilde{N}_{Z}^1,\epsilon)\le-\frac{\sqrt{N}}{2}$, $N_{Z}^1\ge0$. Here, $\tilde{N}_{Z}^1$ denotes the estimated value of $N_{Z}^1$. 

We apply Eq.~(\ref{eq:lineardecoyS1}) to Eq.~(\ref{eq:KatodecoyZ1}), and then using Kato's inequality, 
we derive upper bounds for $\sum_{i=1}^N P(Z^{\U{det},S}_i=1|F_{i-1})$ and 
$\sum_{i=1}^N P(Z^{\U{det},V}_i=1|F_{i-1})$ using their corresponding random variables 
$N_{Z,S}=\sum_{i=1}^N Z^{\U{det},S}_i$ 
and $N_{Z,V}=\sum_{i=1}^N Z^{\U{det},V}_i$, respectively, as well as a lower bound for 
$\sum_{i=1}^N P(Z^{\U{det},D}_i=1|F_{i-1})$ using its corresponding random variable 
$N_{Z,D}=\sum_{i=1}^N Z^{\U{det},D}_i$. This ends the proof of Theorem~\ref{theorem:decoy1ZL}. 

\section{Simulation of key rate}
\label{sec:simulation}

\begin{table}[t]
\centering
\renewcommand{\arraystretch}{1.2}
\begin{tabular}{cccccc}
\hline\hline
$d$ & $\delta_{\U{mis}}$  & $f$ & $\epsilon_c$ & $\epsilon_s$ \\
\hline
$10^{-9}$ (Fig.~\ref{fig:keyrate_dark-9}) &
\multirow{2}{*}{$3\%$} &
\multirow{2}{*}{$1.16$} &
\multirow{2}{*}{$\frac{10^{-10}}{2}$} &
\multirow{2}{*}{$\frac{10^{-10}}{2}$} \\
$10^{-6}$ (Fig.~\ref{fig:keyrate_dark-6}) & & & & & \\
\hline\hline
\end{tabular}
\caption{
List of parameters used in the key-rate simulation. 
Here, $d$ denotes the dark-count probability, 
$\delta_{\U{mis}}$ is the polarization misalignment of the system, 
$f$ represents the error-correction inefficiency, 
$\epsilon_c$ is the correctness parameter, and 
$\epsilon_s$ is the secrecy parameter to satisfy the overall security parameter 
$\epsilon_{\U{sec}}=\epsilon_s + \epsilon_c=10^{-10}$.
}
\label{table:simulation}
\end{table}

\begin{figure*}[t]
    \centering
    \includegraphics[width=11.5cm]{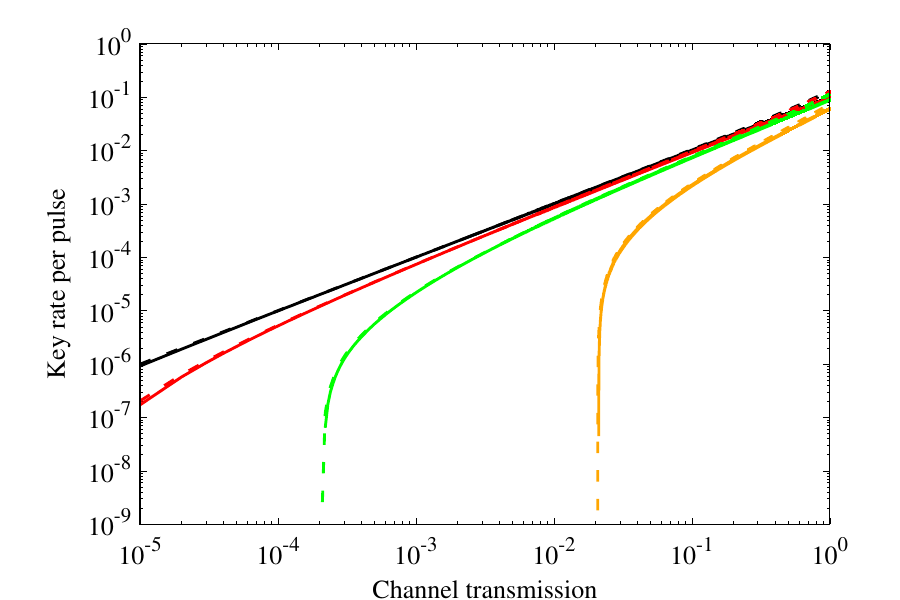}
    \caption{
     Secret key rate $R:=\ell/N$ per single emitted pulse as a function of the overall channel transmission $\eta$, with the dark count probability $d = 10^{-9}$. From bottom to top, we plot the key rates for the number of pulses emitted $N = 10^8$, $10^{10}$, $10^{12}$, and the asymptotic case, using solid lines. For comparison, we also plot the key rates of the active polarization-encoding decoy-state BB84 protocol with dashed lines in the same manner.
    }
\label{fig:keyrate_dark-9}
\end{figure*}

\begin{figure*}[t]
    \centering
    \includegraphics[width=11.5cm]{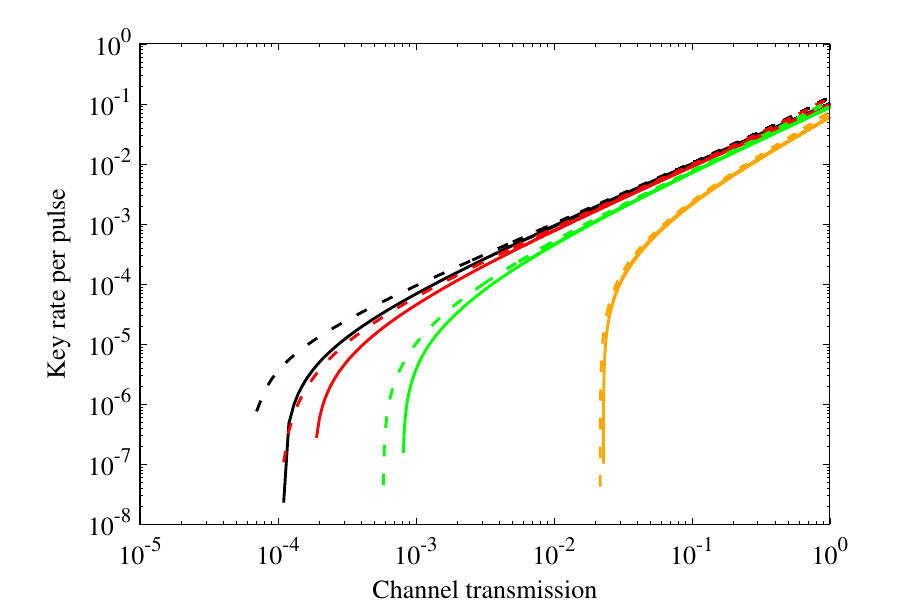}
    \caption{
    Secret key rate $R:=\ell/N$ per single emitted pulse as a function of the overall channel transmission $\eta$, with the dark count probability $d = 10^{-6}$. From bottom to top, we plot the key rates for $N = 10^8$, $10^{10}$, $10^{12}$, and the asymptotic case, using solid lines. For comparison, we also plot the key rates of the active polarization-encoding decoy-state BB84 protocol with dashed lines in the same manner.
    }
\label{fig:keyrate_dark-6}
\end{figure*}

In this section, we present the simulation results of the key rate $R:=\ell/N$ for our passive decoy-state BB84 protocol. 
For comparison, we also show the simulation results for the polarization-encoded decoy-state BB84 protocol employing an active basis choice on Bob's side.

From Eqs.~(\ref{eq:eqell}), (\ref{eq:NPA}), (\ref{eq:omega}) and Theorems~\ref{theorem:Nphupper} and \ref{theorem:decoy1ZL}, the secret key length $\ell$ can be expressed as
\begin{align}
\ell=N_Z^{1,\U{L}}\left[1-h\(\frac{N_{\U{ph}}^{1,\U{U}}}{N_Z^{1,\U{L}}}\)\right]-\xi-N_{\U{EC}}
\label{eq:keylength}
\end{align}
with $\epsilon_{\rm sec}=\epsilon_c+\sqrt{2}\sqrt{\epsilon_{\U{ph}}+2^{-\xi}}$-secure.
From Theorems~\ref{theorem:Nphupper} and \ref{theorem:decoy1ZL}, the failure probability $\epsilon_{\U{ph}}$ in Eq.~(\ref{eq:kphprobability}) for estimating the amount of privacy amplification is given by $9\epsilon$, where $\epsilon$ is the failure probability appearing in these theorems. We set $\epsilon_c = 10^{-10}/2$, $\xi = 71$, and $\epsilon = 10^{-20}/144$ to ensure that the overall security parameter $\epsilon_{\rm sec}$ is $10^{-10}$.

In this simulation, we assume that the quantum channel between Alice and Bob is a linear-loss channel with channel transmission $\eta$, which includes the detection efficiency. 
We also assume that the practical cost of bit-error correction is given by $N_{\mathrm{EC}} = 1.16\, N_{\mathrm{sift}}\, h(e_{\mathrm{bit}})$, where $N_{\mathrm{sift}} = \sum_{\omega \in \{S, D, V\}} N_{Z,\omega}$, $f=1.16$ is the error correction inefficiency \cite{errorefficiency}, and $e_{\mathrm{bit}}$ is the bit error rate. 
To account for the polarization misalignment in the system, we assume a misalignment-induced bit-error rate of 
$\delta_{\U{mis}}=3\%$ in both the $Z$ and $X$ bases. 
All the data that would be obtained in our protocol under this channel model are described in Appendix~\ref{app:simulation}. 
The system parameters used in our simulation are summarized in Table~\ref{table:simulation}.

Under these conditions, we optimize the key rate $R$ with respect to $p_Z = 1 - q>0.5$ and the signal intensity $\mu_S > \mu_V=0.05$ under the constraints $p_S = 0.8$ and $p_D = p_V = 0.1$. The results are shown in Figs.~\ref{fig:keyrate_dark-9} and \ref{fig:keyrate_dark-6}. 
We compare secret-key-generation performance between passive- and active-measurement setups. For reference, we also plot the key rate of a polarization-encoded decoy-state BB84 protocol with active basis choice on Bob's side, evaluated under the same linear-loss channel model and using the security proof in Sec. II.E of Ref.~\cite{pspd}.
As shown in Fig.~\ref{fig:keyrate_dark-9}, when the dark count probability $d$ is $10^{-9}$, the key rates for passive and active measurement setups show no significant difference, regardless of the number of emitted pulses $N$. At short distances (i.e., near $\eta=0$), this gap becomes slightly more pronounced compared to other values of $\eta$. This is due to an increased phase error rate in the passive measurement setup, caused by cross-click events resulting from multiphoton emissions by Alice. Specifically, this effect originates mainly from the first term of $N_{D}^{\U{Cross}}$ in Eq.~(\ref{eq:NDcross}).

As shown in Fig.~\ref{fig:keyrate_dark-6}, when the dark count probability is $10^{-6}$, the difference in key rates between the active and passive measurement setups becomes more pronounced than in Fig.~\ref{fig:keyrate_dark-9}. This indicates an intrinsic sensitivity of passive measurement setups to dark counts \cite{Kawakami2025}. To see this, consider the signal-to-noise (S/N) ratio, where the noise originates from dark counts. In passive setups, a smaller value of $q$ leads to a lower signal level in the $X$ line, while the noise in the $X$ line remains unchanged. In contrast, in active setups, a smaller probability of choosing the $X$ basis does not change both the signal level and noise conditioned on choosing the $X$ basis. Therefore, when $q$ is small (i.e., the basis choice is highly biased), the S/N ratio in passive setups is worse than in active setups. As a result, the key rate for the passive basis choice becomes lower than that for the active basis choice unless the noise (i.e., dark count rate) is negligible compared to the signal level (i.e., detection rate). 

These results indicate that the better the performance of the photon detectors, the smaller the difference between the active and passive setups becomes. For example, when the dark count probability is $d=10^{-9}$~\cite{detectorexperiment} and the detection efficiency $\eta_{\U{det}}$ is 80\%, the red curves in Fig.~\ref{fig:keyrate_dark-9} demonstrate that long-distance QKD over approximately 245~km is feasible with both passive and active measurement setups.

\section{Conclusions}
\label{sec:conclusions}
We have proven the finite-key security of the decoy-state BB84 protocol with passive biased basis choice at the receiver. We have provided a simple, fully analytical security proof that eliminates the need to estimate Bob's single-photon detection rate and have shown that the discrepancy in phase error rates between active and passive implementations can be upper-bounded by the cross-click rate, which is typically negligible. 
Our numerical simulations have shown that the passive protocol attains essentially the same key generation rate as the active protocol while retaining practical implementation advantages. 
While this work has focused on passive measurement by Bob, an important direction for future work is to extend our analysis to the fully passive BB84 protocol~\cite{fullypassive, fullypassivefinite} with passive state preparation by Alice.

\section*{Data availability statement}
No new data were created or analyzed in this study.

\section*{Acknowledgment}
We thank Toshihiko Sasaki, Yuki Takeuchi, Takuya Ikuta, Toshimori Honjo, Lars Kamin and Devashish Tupkary for helpful discussions. A. Mizutani is partially supported by JSPS KAKENHI Grant Number JP24K16977. G. Kato is partially supported by JST CREST Grant Number JPMJCR2113, and JSPS KAKENHI Grant Number JP23K25793.

\appendix
\section{Kato's inequality}
\label{app:kato}
In this section, we explain Kato's inequality~\cite{Katoinequality}, which is employed in the proofs of Theorems~\ref{theorem:Nphupper} and~\ref{theorem:decoy1ZL}.

Let $\xi_0, \xi_1, \ldots, \xi_N$ be a sequence of Bernoulli random variables with $\xi_0=0$, 
and define $\Lambda_l = \sum_{i=1}^{l} \xi_i$. 
Let $\{F_i\}_{i=0}^N$ be a filtration with $F_i$ identifying the random variables including $\{\xi_0,...,\xi_i\}$. 

\subsection{Upper bound on the sum of probabilities}
\label{app:secupper}
Kato's inequality states that for any $a,b$ such that $b\ge|a|$, 
\begin{align}
P\Bigg\{
\sum_{i=1}^NP(\xi_i=1|F_{i-1})\ge \Lambda_N+
\frac{bN+a\left(2\Lambda_N-N\right)}{\sqrt{N}}
\Bigg\}
\le 
\exp\(-\frac{2b^2-2a^2}{(1+\frac{4a}{3\sqrt{N}})^2}\).
\label{eq:katoproupper}
\end{align}
Here, the deviation term in Kato's inequality corresponds to the second term on the right-hand side of the inequality inside the probability $P\{\}$, while the probability on the right-hand side of Eq.~(\ref{eq:katoproupper}) is a constant $\epsilon$ (with $0<\epsilon<1$). 
Under the conditions that this probability is $\epsilon$ and $b\ge|a|$, the values of $a$ and $b$ that minimize the term $\left[bN+a(2C-N)\right]/\sqrt{N}$, for a fixed constant $0\le C\le N$, are given by 
\begin{align}
a_{\U{U}}(N,C,\epsilon):=\frac{216\sqrt{N}C(N-C)\ln\epsilon-48N^{\frac{3}{2}}(\ln\epsilon)^2
+27\sqrt{2}(N-2C)\sqrt{-N^2(\ln\epsilon)[9C(N-C)-2N\ln\epsilon]}}
{4(9N-8\ln\epsilon)[9C(N-C)-2N\ln\epsilon]
}
\label{kato-ineq-def-a}
\end{align}
and
\begin{align}
b_{\U{U}}(N,C,\epsilon):=
\sqrt{a_{\U{U}}(N,C,\epsilon)^2+\frac{1}{2}\left(1+\frac{4a_{\U{U}}(N,C,\epsilon)}{\sqrt{9N}}\right)^2\ln\frac{1}{\epsilon}},
\label{kato-ineq-def-b}
\end{align}
respectively~\cite{tightnpj}. 
The important point here is that $\Lambda_N$ cannot be identified with $C$, since $\Lambda_N$ is a random variable rather than a constant. 
Therefore, we use an estimated value of $\Lambda_N$, which we denote by $\tilde{\Lambda}_N$. 
We then adopt $a = a_{\U{U}}(N,\tilde{\Lambda}_N,\epsilon)$ and $b =b_{\U{U}}(N,\tilde{\Lambda}_N,\epsilon)$, 
and by substituting these into Eq.~(\ref{eq:katoproupper}), 
we find that 
\begin{align}
\sum_{i=1}^NP(\xi_i=1|F_{i-1})\le \Lambda_N+
\Delta_{\U{U}}(\Lambda_N,\tilde{\Lambda}_N,N,\epsilon)
\label{eq:resultkatoU}
\end{align}
holds except with probability $\epsilon$. Here, the deviation term $\Delta_{\U{U}}(\Lambda_N,\tilde{\Lambda}_N,N,\epsilon)$ to upper-bound the sum of probabilities 
is defined in Eq.~(\ref{eq:deviationfunc}).

The derivation of  Eqs.~(\ref{kato-ineq-def-a}) and (\ref{kato-ineq-def-b}) proceeds as follows.
By solving the constraint 
\[
\epsilon=\exp\(-\frac{2b^2-2a^2}{(1+\frac{4a}{3\sqrt{N}})^2}\)
\]
for $b$, we obtain 
\begin{align}
b=b(a):=
\sqrt{a^2+\frac{1}{2}\left(1+\frac{4a}{\sqrt{9N}}\right)^2\ln\frac{1}{\epsilon}}.
\label{app:b}
\end{align}
Substituting this expression into the numerator $\phi(a):=b(a)N+a\left(2C-N\right)$ 
of the deviation term, we find that $\phi(a)$ is a convex function of $a$, since $\phi''(a)>0$ holds for any $a\in\mathbb{R}$. 
Therefore, differentiating $\phi(a)$ with respect to $a$ and solving $\phi'(a)=0$ gives a unique stationary point that minimizes the deviation term. A direct calculation shows that Eq.~(\ref{kato-ineq-def-a}) satisfies $\phi'(a)=0$, and substituting $a=a_{\U{U}}(N,C,\epsilon)$ into Eq.~(\ref{app:b}) gives Eq.~(\ref{kato-ineq-def-b}).

\subsection{Lower bound on the sum of probabilities}
By replacing $\xi_i\to 1-\xi_i$ and $a\to-a$ in Eq.~(\ref{eq:katoproupper}), we also have
\begin{align}
P\Bigg\{
\Lambda_N\ge
\sum_{i=1}^NP(\xi_i=1|F_{i-1}) +
\frac{bN+a\left(2\Lambda_N-N\right)}{\sqrt{N}}
\Bigg\}
\le 
\exp\(-\frac{2b^2-2a^2}{(1-\frac{4a}{3\sqrt{N}})^2}\).
\label{eq:katoprolower}
\end{align}
We repeat a discussion similar to that in Sec.~\ref{app:secupper}. 
That is, when the right-hand side of Eq.~(\ref{eq:katoprolower}) is $\epsilon$, the optimal values of $a$ and $b$ 
that minimize the deviation term $\left[bN+a(2C-N)\right]/\sqrt{N}$ are obtained as follows.
\begin{align}
a_{\U{L}}(N,C,\epsilon):=\frac{-216\sqrt{N}C(N-C)\ln\epsilon+48N^{\frac{3}{2}}(\ln\epsilon)^2
+27\sqrt{2}(N-2C)\sqrt{-N^2(\ln\epsilon)[9C(N-C)-2N\ln\epsilon]}}
{4(9N-8\ln\epsilon)[9C(N-C)-2N\ln\epsilon]
}
\label{kato-ineq-def-a1}
\end{align}
\begin{align}
b_{\U{L}}(N,C,\epsilon):=
\sqrt{a_{\U{L}}(N,C,\epsilon)^2+\frac{1}{2}\left(1-\frac{4a_{\U{L}}(N,C,\epsilon)}{\sqrt{9N}}\right)^2\ln\frac{1}{\epsilon}}
\label{kato-ineq-def-b2}
\end{align}
By adopting $a = a_{\U{L}}(N,\tilde{\Lambda}_N,\epsilon)$ and $b =b_{\U{L}}(N,\tilde{\Lambda}_N,\epsilon)$, 
and by substituting these into Eq.~(\ref{eq:katoprolower}), we find that 
\begin{align}
\sum_{i=1}^NP(\xi_i=1|F_{i-1})\ge
\Lambda_N-\Delta_{\U{L}}(\Lambda_N,\tilde{\Lambda}_N,N,\epsilon)
\label{eq:resultkatoL}
\end{align}
holds except with probability $\epsilon$. Here, the deviation term $\Delta_{\U{L}}(\Lambda_N,\tilde{\Lambda}_N,N,\epsilon)$ to lower-bound the sum of probabilities 
is defined in Eq.~(\ref{eq:deviationfunc2}).

Note that the derivations of $a_{\U{L}}(N,C,\epsilon)$ and $b_{\U{L}}(N,C,\epsilon)$ follow the same reasoning as 
explained in the last paragraph of Sec.~\ref{app:secupper}.

\section{Proof of Lemma~\ref{theorem1:probability}}
\label{app:proofLemma1}
In this appendix, we prove Lemma~\ref{theorem1:probability}. 
From Eq.~(\ref{eq:Enphi}), we have
\begin{align}
&P(E^{\U{ph},1}_i=1|F_{i-1})=
P\(\alpha_i=Z,n_i=1|F_{i-1}\)P\(\beta_i=Z,e_{X,i}=1|\alpha_i=Z,n_i=1,F_{i-1}\)
\notag\\
=&p_Zp^{\U{int}}(1)\sum_{m=0}^{\infty}\underbrace{P(m_i=m|\alpha_i=Z,n_i=1,F_{i-1})}_{=:p_m}
\underbrace{P(\beta_i=Z,e_{X,i}=1|m_i=m,\alpha_i=Z,n_i=1,F_{i-1})}_{=:r_m}.
\end{align}
The first equation follows from applying the Bayes rule. 
The second equation is obtained by marginalizing over the number of photons $m_i$ entering Bob's measurement setup 
and applying the Bayes rule.

When the vacuum state is input to Bob's measurement setup (namely, $m_i=0$), Bob obtains a click event originating from the 
dark count. Therefore, conditioned on $m_i=0$, the probability $r_0$ that $\beta_i=Z$ and a phase error event occurs 
(namely, $e_{X,i}=1$) is given by $r_0=(1-d)^2[1-(1-d)^2]/2$. 
Note that $(1-d)^2$ represents the probability that no dark count occurs in the two detectors on the $X$ line, and 
the factor $1/2$ in $r_0$ reflects the fact that a phase error event occurs randomly when a detection is caused by a dark count. 
Also, $r_m$ for $m\ge1$ is calculated as
\begin{align}
r_m&=P(\beta_i=Z|m_i=m,\alpha_i=Z,n_i=1,F_{i-1})
\underbrace{P(e_{X,i}=1|\beta_i=Z,m_i=m,\alpha_i=Z,n_i=1,F_{i-1})}_{=:e_m}
\notag\\
&=(1-q)^m(1-d)^2e_m.
\end{align}
The first equation follows from applying the Bayes rule. 
The second equation follows from the fact that, conditioned on Bob receiving $m$ photons immediately before his measurement setup, he obtains $\beta_i=Z$ when all $m$ photons go to the $Z$ line and no dark count occurs in the two detectors on the $X$ line, 
and from the fact that the action of Bob's beam splitter depends only on the input photon number, as stated in Assumption~\ref{Ass:BobBS}. 

Hence, $P(E^{\U{ph},1}_i=1|F_{i-1})$ is rewritten as
\begin{align}
P(E^{\U{ph},1}_i=1|F_{i-1})=p_Zp^{\U{int}}(1)
(1-d)^2
\left[p_0\frac{1-(1-d)^2}{2}+\sum_{m=1}^{\infty}p_m(1-q)^me_m\right].
\label{eq:PEPH1}
\end{align}
By performing a similar calculation to obtain Eq.~(\ref{eq:PEPH1}) for $P(E^{X,1}_i=1|F_{i-1})$ defined in 
Eq.~(\ref{eq:EXNEXOMEGA}), we obtain
\begin{align}
P(E^{X,1}_i=1|F_{i-1})=p_Xp^{\U{int}}(1)(1-d)^2
\left[p_0\frac{1-(1-d)^2}{2}+\sum_{m=1}^{\infty}p_mq^me_m\right].
\label{eq:PEX1}
\end{align}
Here, we use the fact that the single-photon state emitted by Alice is basis independent, namely, $\sum_{a_i\in\{0,1\}}\ket{{1_{\phi(a_i,Z)}}}\bra{{1_{\phi(a_i,Z)}}}_{S_i}/2=\sum_{a_i\in\{0,1\}}\ket{{1_{\phi(a_i,X)}}}\bra{{1_{\phi(a_i,X)}}}_{S_i}/2$. Consequently, 
$P(m_i=m|\alpha_i=X,n_i=1,F_{i-1})=p_m$ and 
$P(e_{X,i}=1|\beta_i=X,m_i=m,\alpha_i=X,n_i=1,F_{i-1})=e_m$. 

Then, using Eqs.~(\ref{eq:PEPH1}) and (\ref{eq:PEX1}) leads to
\begin{align}
&
\frac{1}{p^{\U{int}}(1)}\(P(E^{\U{ph},1}_i=1|F_{i-1})-\frac{p_Z(1-q)}{p_Xq}P(E^{X,1}_i=1|F_{i-1})\)
\notag\\
=&p_Z(1-d)^2\Bigg\{p_0\frac{1-(1-d)^2}{2}\frac{2q-1}{q}
+(1-q)\sum_{m=2}^{\infty}p_me_m[(1-q)^{m-1}-q^{m-1}]\Bigg\}\notag\\
\le&
p_Z(1-d)^2\Bigg\{
p_0\frac{1-(1-d)^2}{2}\frac{2q-1}{q}
+(1-q)\sum_{m=2}^{\infty}p_m[(1-q)^{m-1}-q^{m-1}]\Bigg\},
\label{eq:linearityph}
\end{align}
where the inequality follows by $e_m\le1$ for $m\ge2$ with Assumption~\ref{Ass:BobBS}. 
We further upper-bound the right-hand-side of Eq.~(\ref{eq:linearityph}) using the probability 
that the cross-click event occurs. For this, from Eq.~(\ref{eq:CnComega}), the probability of obtaining 
this event is written as
\begin{align}
\frac{
P(C^{1}_i=1|F_{i-1})}{p^{\U{int}}(1)}=
p_0[1-(1-d)^2]^2+\sum_{m=1}^{\infty}p_m\left[1-(q^m+(1-q)^m)(1-d)^2\right].
\label{eq:PC1i}
\end{align}
Let $T:=p_Z(1-d)^2(1-q)(1-2q)>0$ denote the prefactor of $p_{m=2}$ in Eq.~(\ref{eq:linearityph}), and 
$S:=1-[q^2+(1-q)^2](1-d)^2>0$ denote the prefactor of $p_{m=2}$ in Eq.~(\ref{eq:PC1i}). Then, combining  
Eqs.~(\ref{eq:linearityph}) and (\ref{eq:PC1i}) gives
\begin{align}
&
\frac{1}{p^{\U{int}}(1)}\(P(E^{\U{ph},1}_i=1|F_{i-1})-\frac{p_Z(1-q)}{p_Xq}P(E^{X,1}_i=1|F_{i-1})-\frac{T}{S}P(C^{1}_i=1|F_{i-1})\)
\notag\\
\le&p_0\left\{-p_Z(1-d)^2\frac{1-(1-d)^2}{2}\frac{1-2q}{q}-\frac{T}{S}[1-(1-d)^2]^2\right\}
-p_1\frac{T}{S}[1-(1-d)^2]\notag\\
+&p_Z(1-d)^2(1-q)\sum_{m\ge3}p_m[(1-q)^{m-1}-q^{m-1}]-\frac{T}{S}\sum_{m\ge3}p_m[1-(q^m+(1-q)^m)(1-d)^2]
\notag\\
\le&\sum_{m\ge3}p_m
\left\{
\underbrace{
p_Z(1-d)^2(1-q)[(1-q)^{m-1}-q^{m-1}]
}_{=:f_m}
-
\underbrace{
\frac{T}{S}[1-(q^m+(1-q)^m)(1-d)^2]
}_{=:g_m}
\right\}
\notag\\
\le&0.
\label{eq:linearityph2}
\end{align}
The second inequality follows because the coefficients of $p_0$ and $p_1$ are negative for $q<1/2$. 
The third inequality follows from the fact that $f_m$ and $g_m$ are functions that are monotonically decreasing and monotonically increasing for $m\ge3$, respectively, and hence $f_m-g_m\le f_3-g_3<0$. Therefore, we have
\begin{align}
P(E^{\U{ph},1}_i=1|F_{i-1})
\le c_1P(E^{X,1}_i=1|F_{i-1})+c_2P(C^{1}_i=1|F_{i-1}),
\label{eq:resultlinear}
\end{align}
with $c_1$ and $c_2$ defined in Eq.~(\ref{eq:defc1c2}). 

Using the decoy-state method~\cite{Ma2005}, the two probabilities appearing in the upper bound in Eq.~(\ref{eq:resultlinear}) 
are given as follows:
\begin{align}
P(E^{X,1}_i=1|F_{i-1})
\le
\frac{p^{\U{int}}(1)}{\mu_D}\(\frac{e^{\mu_D}}{p_D}P(E^{X,D}_i=1|F_{i-1})-\frac{P(E^{X,V}_i=1|F_{i-1})}{p_V}\)
\end{align}
\begin{align}
P(C^{1}_i=1|F_{i-1})
\le
\frac{p^{\U{int}}(1)}{\mu_D}\(\frac{e^{\mu_D}}{p_D}P(C^{D}_i=1|F_{i-1})-\frac{P(C^{V}_i=1|F_{i-1})}{p_V}\).
\end{align}
Substituting these bounds into Eq.~(\ref{eq:resultlinear}) 
results in Eq.~(\ref{eq:RHSlemma}), which ends the proof of Lemma~\ref{theorem1:probability}.

\section{Application of Kato's inequality in the proof of Theorem~\ref{theorem:Nphupper}}
\label{app:applKato}
In this appendix, we evaluate upper bounds on the sums of the remaining four probabilities obtained by summing Eq.~(\ref{eq:RHSlemma}) over $i=1,...,N$. Specifically, we upper-bound $\sum_{i=1}^N P(C^{D}_i=1|F_{i-1})$ 
and lower-bound $\sum_{i=1}^NP(E^{\U{ph},1}_i=1|F_{i-1})$, $\sum_{i=1}^N P(E^{X,V}_i=1|F_{i-1})$ and $\sum_{i=1}^N P(C^{V}_i=1|F_{i-1})$ with their corresponding random variables using Kato's inequality. In the description of these bounds, 
$\tilde{N}_D^{\U{Cross}}$, $\tilde{N}_{X,V}^{\U{Error}}$, 
$\tilde{N}_V^{\U{Cross}}$ and $\tilde{N}_{\U{ph}}^1$ denote estimated values of $N_{D}^{\U{Cross}}$, $N_{X,V}^{\U{Error}}$, 
$N_{V}^{\U{Cross}}$ and $N_{\U{ph}}^1$, respectively. 
\\\\
{\bf 1: Upper bound on $\sum_{i=1}^N P(C^{D}_i=1|F_{i-1})$}
\\
From Kato's inequality in Eq.~(\ref{eq:resultkatoU}), the following inequality holds except with probability 
$\epsilon$:
\begin{align}
\sum_{i=1}^N P(C^{D}_i=1|F_{i-1})
&\le N_{D}^{\U{Cross}}+\Delta_{\U{U}}(N_{D}^{\U{Cross}},\tilde{N}_D^{\U{Cross}},N,\epsilon).
\label{eq:KatoCD}
\end{align}
\\
{\bf 2: Lower bounds on $\sum_{i=1}^N P(E^{X,V}_i=1|F_{i-1})$ and $\sum_{i=1}^N P(C^{V}_i=1|F_{i-1})$}
\\
From Kato's inequality in Eq.~(\ref{eq:resultkatoL}), each of the following inequalities holds except with probability $\epsilon$: 
\begin{align}
\sum_{i=1}^NP(E^{X,V}_i=1|F_{i-1})&\ge 
N_{X,V}^{\U{Error}}-
\Delta_{\U{L}}(N_{X,V}^{\U{Error}},\tilde{N}_{X,V}^{\U{Error}},N,\epsilon),
\label{eq:NXerrorV}
\\
\sum_{i=1}^N P(C^{V}_i=1|F_{i-1})
&\ge 
N_{V}^{\U{Cross}}-
\Delta_{\U{L}}(N_{V}^{\U{Cross}},\tilde{N}_V^{\U{Cross}},N,\epsilon).
\label{eq:NcrossV}
\end{align}
\\
{\bf 3: Lower bound on $\sum_{i=1}^N P(E^{\U{ph},1}_i=1|F_{i-1})$}
\\
From Kato's inequality in Eq.~(\ref{eq:katoprolower}), 
\begin{align}
N_{\U{ph}}^1\le\sum_{i=1}^NP(E^{\U{ph},1}_i=1|F_{i-1})+
\frac{bN+a(2N_{\U{ph}}^1-N)}{\sqrt{N}}
\label{eq:NphKato}
\end{align}
holds except with probability $\epsilon:=\exp\left[-\frac{2b^2-2a^2}{(1-\frac{4a}{3\sqrt{N}})^2}\right]$ 
for any $a,b\in \mathbb{R}$ satisfying $b\ge|a|$. 

If $a^{\U{ph}}:=a_{\U{L}}(N,\tilde{N}_{\U{ph}}^1,\epsilon)<\frac{\sqrt{N}}{2}$, by setting $a$ and $b$ as $a^{\U{ph}}$ and $b^{\U{ph}}:=b_{\U{L}}(N,\tilde{N}_{\U{ph}}^1,\epsilon)$, respectively, Eq.~(\ref{eq:NphKato}) implies that
\begin{align}
N_{\U{ph}}^1\le
\left(1-\frac{2a^{\U{ph}}}{\sqrt{N}}\right)^{-1}
\left[\sum_{i=1}^NP(E^{\U{ph},1}_i=1|F_{i-1})+(b^{\U{ph}}-a^{\U{ph}})\sqrt{N}
\right]
\end{align}
holds except with probability $\epsilon$. On the other hand, if $a_{\U{L}}(N,\tilde{N}_{\U{ph}}^1,\epsilon)\ge\frac{\sqrt{N}}{2}$, we have $N_{\U{ph}}^1\le N$. 

Combining this inequality with Eqs.~(\ref{eq:RHSlemma}), (\ref{eq:katoNXDerror}), and (\ref{eq:KatoCD})-(\ref{eq:NcrossV}) gives that
\begin{align}
&N_{\U{ph}}^1
\le 
\left(1-\frac{2a^{\U{ph}}}{\sqrt{N}}\right)^{-1}
\notag\\
&\Bigg\{
\frac{p^{\U{int}}(1)}{\mu_D}\Bigg[c_1\(\frac{e^{\mu_D}
(N_{X,D}^{\U{Error}}+\Delta_{\U{U}}(N_{X,D}^{\U{Error}},\tilde{N}_{X,D}^{\U{Error}},N,\epsilon))
}{p_D}
-\frac{
N_{X,V}^{\U{Error}}-\Delta_{\U{L}}(N_{X,V}^{\U{Error}},\tilde{N}_{X,V}^{\U{Error}},N,\epsilon)
}{p_V}\)
\notag\\
&+c_2\(\frac{e^{\mu_D}
(N_{D}^{\U{Cross}}+\Delta_{\U{U}}(N_{D}^{\U{Cross}},\tilde{N}_D^{\U{Cross}},N,\epsilon))
}{p_D}
-\frac{
N_{V}^{\U{Cross}}-\Delta_{\U{L}}(N_{V}^{\U{Cross}},\tilde{N}_V^{\U{Cross}},N,\epsilon)
}{p_V}\)\Bigg]
\notag\\
&
+(b^{\U{ph}}-a^{\U{ph}})\sqrt{N}
\Bigg\}
\end{align}
holds except with probability $5\epsilon$. This ends the proof of Theorem~\ref{theorem:Nphupper}.

\section{Simulated data under the linear-loss channel model}
\label{app:simulation}
In this appendix, we calculate the simulated data used for the key-rate simulation presented in Sec.~\ref{sec:simulation}, 
under the linear-loss channel model with channel transmission $\eta$. 
To evaluate the secret key length $\ell$ in Eq.~(\ref{eq:keylength}), we need to derive the following quantities: 
$N_{Z,\omega}$ for $N^{1,\U{L}}_{Z}$, the bit error rate $e_{\U{bit}}$ for $N_{\U{EC}}$, and 
$N^{\U{Cross}}_{\omega}$ and $N^{\U{Error}}_{X,\omega}$ for $N^{1,\U{U}}_{\U{ph}}$, 
each of which is calculated in the following subsections.

\subsection{Calculation of $N_{Z,\omega}$}
As defined in Sec.~\ref{sec:actualprotocol}, $N_{Z,\omega}$ denotes the number of events in which Alice chooses the $Z$ basis and the intensity label $\omega\in\{S,D,V\}$, and Bob obtains a click event in the $Z$ basis. 
For the purpose of simulation, we assume that $N_{Z,\omega}$ is given by
\begin{align}
N_{Z,\omega} = N p_Z p_{\omega} \, p(\beta_i = Z \mid \alpha_i = Z, \omega_i = \omega),
\label{appe:nzo}
\end{align}
where $p(\beta_i = Z \mid \alpha_i = Z, \omega_i = \omega)$ represents the probability that 
Bob obtains a click event in the $Z$ basis conditioned on Alice choosing the $Z$ basis and the intensity label $\omega$. 
This probability is expressed as
\begin{align}
p(\beta_i=Z|\alpha_i=Z,\omega_i=\omega)=&e^{-\mu_{\omega}\eta q}(1-d)^2
\left[1-e^{-\mu_{\omega}\eta(1-q)}(1-d)^2\right].
\label{appe:pzo}
\end{align}
Here, 
the term $e^{-\mu_{\omega}\eta q}(1-d)^2$ represents the probability that the optical pulse in the $X$ line is in the vacuum state 
and no dark count occurs in the $X$ line. 
In contrast, the probability inside the square brackets represents the probability that a detection event occurs in the $Z$ line, 
either because an optical pulse arrives or due to a dark count.

\subsection{Calculation of $e_{\U{bit}}$}
Here, $e_{\U{bit}}$ denotes the bit error rate in the sifted key. For the purpose of simulation, we assume a misalignment error of 
$\delta_{\U{mis}}=3\%$ (as listed in Table~\ref{table:simulation}), and $e_{\U{bit}}$ is expressed as
\begin{align}
e_{\U{bit}} = \frac{N_{\U{sift}}^{\U{error}}}{N_{\U{sift}}} + \delta_{\U{mis}}
\end{align}
with
\begin{align}
N_{\rm sift}^{\rm error}=\sum_{\omega \in \{S, D, V\}}Np_{\omega}p_Z \, p(\U{Error},\beta_i=Z|\alpha_i=Z,\omega_i=\omega).
\end{align}
Here, $N_{\U{sift}}$ denotes the length of the sifted key, as defined in Sec.~\ref{sec:actualprotocol}. 
The term $N_{\U{sift}}^{\U{error}}$ represents the number of sifted bits in which Alice and Bob 
obtain different bit values in the $Z$ basis, and $p(\U{Error},\beta_i=Z|\alpha_i=Z,\omega_i=\omega)$ denotes the probability that Bob obtains a click event in the $Z$ basis 
and that Alice and Bob share a bit error in the $Z$ basis, 
conditioned on Alice choosing the $Z$ basis and the intensity label $\omega$. 
This probability is expressed as
\begin{align}
p(\U{Error},\beta_i=Z|\alpha_i=Z,\omega_i=\omega)=
e^{-\mu_{\omega}\eta q}(1-d)^2\times 
\left(\underbrace{
e^{-\mu_{\omega}\eta (1-q)}[d(1-d)+d^2/2]}_{\U{term}~A}+
\underbrace{(1-e^{-\mu_{\omega}\eta(1-q)})d/2}_{\U{term}~B}\right).
\end{align}
Here, 
the term $e^{-\mu_{\omega}\eta q}(1-d)^2$ represents the probability that the optical pulse in the $X$ line is in the vacuum state 
and no dark count occurs in the $X$ line. 
The term $A$ denotes the probability that no optical pulse arrives at the $Z$ line and a bit error occurs, either because a dark-count click happens at the error-side detector or because both detectors click due to dark counts followed by random bit assignment. 
The term $B$ represents the probability that an optical pulse arrives at the $Z$ line and a click occurs at the other detector due to a dark count, resulting in a bit error after random bit assignment.

\subsection{Calculation of $N_{X,\omega}^{\U{Error}}$}
As defined in Sec.~\ref{sec:actualprotocol}, $N_{X,\omega}^{\U{Error}}$ denotes the number of events in which 
Alice and Bob obtain different bit values in the $X$ basis when Alice chooses the intensity label $\omega\in\{S,D,V\}$. 
For the purpose of simulation, we assume that $N^{\U{Error}}_{X,\omega}$ is given by
\begin{align}
N^{\U{Error}}_{X,\omega}=N
p_{\omega}p_X \, p(\U{Error},\beta_i=X|\alpha_i=X,\omega_i=\omega)+\delta_{\U{mis}}N_{X,\omega}
\label{NerrorX}
\end{align}
with
\begin{align}
N_{X,\omega} = N p_X p_{\omega} \, p(\beta_i = X \mid \alpha_i = X, \omega_i = \omega).
\end{align}
Here, we assume a misalignment error of $\delta_{\U{mis}}=3\%$ (as listed in Table~\ref{table:simulation}). 
Also, $p(\beta_i = X \mid \alpha_i = X, \omega_i = \omega)$ represents the probability that 
Bob obtains a click event in the $X$ basis conditioned on Alice choosing the $X$ basis and the intensity label $\omega$. 
This probability can be obtained by replacing $q$ with $1-q$ in Eq.~(\ref{appe:pzo}): 
\begin{align}
p(\beta_i=X|\alpha_i=X,\omega_i=\omega)=&e^{-\mu_{\omega}\eta(1-q)}(1-d)^2
\left[1-e^{-\mu_{\omega}\eta q}(1-d)^2\right].
\end{align}
The quantity $p(\U{Error},\beta_i=X|\alpha_i=X,\omega_i=\omega)$ denotes the probability that Bob obtains a click event in the 
$X$ basis and that Alice and Bob share a bit error in the $X$ basis, conditioned on Alice choosing the $X$ basis and the intensity label $\omega$, as expressed in
\begin{align}
p(\U{Error},\beta_i=X|\alpha_i=X,\omega_i=\omega)=
e^{-\mu_{\omega}\eta(1-q)}(1-d)^2\times 
\left(
\underbrace{
e^{-\mu_{\omega}\eta q}[d(1-d)+d^2/2]}_{\U{term}~A}+\underbrace{(1-e^{-\mu_{\omega}\eta q})d/2}_{\U{term}~B}
\right).
\end{align}
Here, 
the term $e^{-\mu_{\omega}\eta(1-q)}(1-d)^2$ represents the probability that the optical pulse in the $Z$ line is in the vacuum state 
and no dark count occurs in the $Z$ line. 
The term $A$ denotes the probability that no optical pulse arrives at the $X$ line and a bit error occurs, either because a dark-count click happens at the error-side detector or because both detectors click due to dark counts followed by random bit assignment. 
The term $B$ represents the probability that an optical pulse arrives at the $X$ line and a click occurs at the other detector due to a dark count, resulting in a bit error after random bit assignment.

\subsection{Calculation of $N_{\omega}^{\U{Cross}}$}
As defined in Sec.~\ref{sec:actualprotocol}, $N_{\omega}^{\U{Cross}}$ denotes the number of cross-click events in which Alice chooses the intensity label $\omega\in\{S,D,V\}$. 
For the purpose of simulation, we assume that $N_{\omega}^{\U{Cross}}$ is given by
\begin{align}
N_{\omega}^{\U{Cross}}= N p_{\omega} \, p_{\U{Cross}|\omega},
\end{align}
where $p_{\U{Cross}|\omega}$ represents the probability that 
Bob obtains a cross-click event conditioned on Alice choosing the intensity label $\omega$. 
This probability is expressed as
\begin{align}
&p_{\U{Cross}|\omega}=(1-e^{-\mu_{\omega}\eta(1-q)})(1-e^{-\mu_{\omega}\eta q})
+(1-e^{-\mu_{\omega}\eta(1-q)})e^{-\mu_{\omega}\eta q}[1-(1-d)^2]\notag\\
&+(1-e^{-\mu_{\omega}\eta q})e^{-\mu_{\omega}\eta(1-q)}[1-(1-d)^2]+e^{-\mu_{\omega}\eta}[1-(1-d)^2]^2.
\label{crossomega}
\end{align}
Here, the first term represents the probability that both the $Z$ and $X$ lines contain one or more photons. 
The second term represents the probability that the $Z$ line contains one or more photons while the $X$ line is in the vacuum state, and at least one detector in the $X$ line clicks due to dark counts. 
The third term represents the probability that the $X$ line contains one or more photons while the $Z$ line is in the vacuum state, and at least one detector in the $Z$ line clicks due to dark counts. 
The fourth term represents the probability that both the $Z$ and $X$ lines are in the vacuum state, while detector clicks occur in both lines due to dark counts. Therefore, we have
\begin{align}
N_{\omega}^{\U{Cross}}=Np_\omega[ 
&(1 - e^{-\mu_{\omega} \eta (1 - q)}) (1 - e^{-\mu_{\omega} \eta q}) 
+ (1 - e^{-\mu_{\omega} \eta (1 - q)}) e^{-\mu_{\omega} \eta q} (1 - (1 - d)^2)
\notag\\
+& (1 - e^{-\mu_{\omega} \eta q}) e^{-\mu_{\omega} \eta (1 - q)} (1 - (1 - d)^2)
+ e^{-\mu_{\omega} \eta} \left(1 - (1 - d)^2\right)^2].
\label{eq:NDcross}
\end{align}

\end{document}